# Physical Conditions of the Ionized Superwind in NGC 253 with VLT/MUSE

Serena A. Cronin[1], Alberto D. Bolatto[1,2], Enrico Congiu[3], Keaton Donaghue[1], Kathryn Kreckel[4], Adam K. Leroy[5,6],
Rebecca C. Levy[7], Sylvain Veilleux[1], Fabian Walter[8], and Lenin Nolasco[1,9]

[1] Department of Astronomy, University of Maryland, College Park, MD 20742, USA; cronin@umd.edu
[2] Joint Space-Science Institute, University of Maryland, College Park, MD 20742, USA
[3] European Southern Observatory (ESO), Alonso de Córdova 3107, Casilla 19, Santiago 19001, Chile
[4] Astronomisches Rechen-Institut, Zentrum für Astronomie der Universität Heidelberg, Mönchhofstraße 12-14, 69120 Heidelberg, Germany
[5] Department of Astronomy, The Ohio State University, 140 West 18th Avenue, Columbus, OH 43210, USA
[6] Center for Cosmology and Astroparticle Physics, 191 West Woodruff Avenue, Columbus, OH 43210, USA
[7] Space Telescope Science Institute, 3700 San Martin Drive, Baltimore, MD 21218, USA
[8] Max-Planck-Institut für Astronomie, Königstuhl 17, 69117 Heidelberg, Germany
[9] Department of Physics and Astronomy, Stony Brook University, Stony Brook, NY 11794-3800, USA
Received 2025 March 14; revised 2025 May 8; accepted 2025 May 8; published 2025 July 1

## Abstract

We present an analysis of the H$\alpha$-emitting ionized gas in the warm phase of the NGC 253 outflow using integral field spectroscopy from the Multi Unit Spectroscopic Explorer. In each spaxel, we decompose H$\alpha$, [N II], and [S II] emission lines into a system of up to three Gaussian components, accounting for the velocity contributions due to the disk and both intercepted walls of an outflow cone. In the approaching southern lobe of the outflow, we find maximum deprojected outflow velocities down to $\sim -500$ km s$^{-1}$. Velocity gradients of this outflowing gas range from $\sim -350$ to $-550$ km s$^{-1}$ kpc$^{-1}$ with increasing distance from the nucleus. Additionally, [N II]/H$\alpha$ and [S II]/H$\alpha$ integrated line ratios are suggestive of shocks as the dominant ionization source throughout the wind. Electron densities, inferred from the [S II] doublet, peak at 2100 cm$^{-3}$ near the nucleus and reach $\lesssim 50$ cm$^{-3}$ in the wind. Finally, at an uncertainty of 0.3 dex on the inferred mass of $4 \times 10^5 \, M_\odot$, the mass-outflow rate of the H$\alpha$-emitting gas in the southern outflow lobe is $\sim 0.4 \, M_\odot$ yr$^{-1}$. This yields a mass-loading factor of $\eta \sim 0.1$ and a $\sim 2$% starburst energy efficiency.

*Unified Astronomy Thesaurus concepts:* Starburst galaxies (1570); Galactic winds (572)

## 1. Introduction

Galactic winds, or outflows, are key players in the evolution of a galaxy from an active spiral to a quiescent elliptical (e.g., S. Veilleux et al. 2005, 2020; D. Colombo et al. 2020; S. L. Ellison et al. 2021). Starburst-driven outflows arise via stellar feedback from the formation of compact, massive star clusters and supernovae. These processes pump energy and momentum into an expanding bubble of hot ($T \sim 10^7$ K) gas that sweeps up the surrounding interstellar medium (ISM) into a shell of shocked material. This shell can be modeled as a bicone exhibiting strong, double-peaked line emission that accounts for both the outflow and the rotation of the galaxy. As it accelerates, the bubble reaches a "blowout" phase where the shocked gas injects its environment with energy and material. In the absence of other drag forces, material moving faster than the escape speed of the galaxy will enrich the surrounding circumgalactic medium with gas, dust, and metals. Material that is unable to escape will accrete back onto the galaxy in a cold "fountain" that may actually spark more star formation (T. M. Heckman et al. 1990; S. Veilleux et al. 2005; B. D. Oppenheimer et al. 2010; S. Veilleux et al. 2020). Galactic winds, as described here, function as "ejective" feedback, where star formation fuel is removed from the galaxy. Winds also prevent the accretion of halo gas onto a galaxy by inducing turbulence and shocks, a function known as "preventative" feedback (V. Pandya et al. 2020;

T. A. Thompson & T. M. Heckman 2024). Both feedback types are key to stunting the growth of a star-forming galaxy.

NGC 253 is an archetypal starburst galaxy with a circumnuclear star formation rate (SFR) of 2.8 $M_\odot$ yr$^{-1}$ (J. Ott et al. 2005; A. K. Leroy et al. 2015). Gas accretion along the bar orbits triggers shocks, enabling efficient gas collapse and heightened star formation activity in the 200 pc circumnuclear ring. This process manifests in compact, massive "super star clusters" (S. F. Portegies Zwart et al. 2010; A. K. Leroy et al. 2018; R. C. Levy et al. 2022) which eventually fuel the galactic superwind (T. A. D. Paglione et al. 2004; E. A. C. Mills et al. 2021).

At a distance of only $d \sim 3.5$ Mpc (E. Congiu et al. 2025, in preparation; R. Rekola et al. 2005; M. J. B. Newman et al. 2024; S. Okamoto et al. 2024), the physical conditions of the NGC 253 wind can be studied in incredible detail, and has been across the electromagnetic spectrum (e.g., M. H. Ulrich 1978; K. Matsubayashi et al. 2009; M. S. Westmoquette et al. 2011; A. D. Bolatto et al. 2013; F. Walter et al. 2017; L. K. Zschaechner et al. 2018; G. I. Günthardt et al. 2019; N. Krieger et al. 2019; S. Lopez et al. 2023). As illustrated in Figure 1(a), the southern lobe of the outflow is pointed toward the observer, while the northern lobe is receding behind the disk due to the galaxy's inclination of 78° (M. S. Westmoquette et al. 2011). Because of this geometry, the northern outflow suffers from heavy obscuration at optical wavelengths caused by the intervening disk. Indeed, H$\alpha$-emitting gas at $T \approx 5000-10^4$ K has been difficult to detect in the northern lobe (M. S. Westmoquette et al. 2011), while both lobes have been detected in the X-ray ($T \approx 10^7$ K; D. K. Strickland et al. 2002; S. Lopez et al. 2023) and the radio (molecular gas of $T \approx 10^2$ K; A. D. Bolatto et al. 2013; F. Walter et al. 2017). With these geometry considerations, the southern outflow in NGC 253 presents an ideal laboratory to compare the properties of Hα-emitting gas with other phases of a superwind.

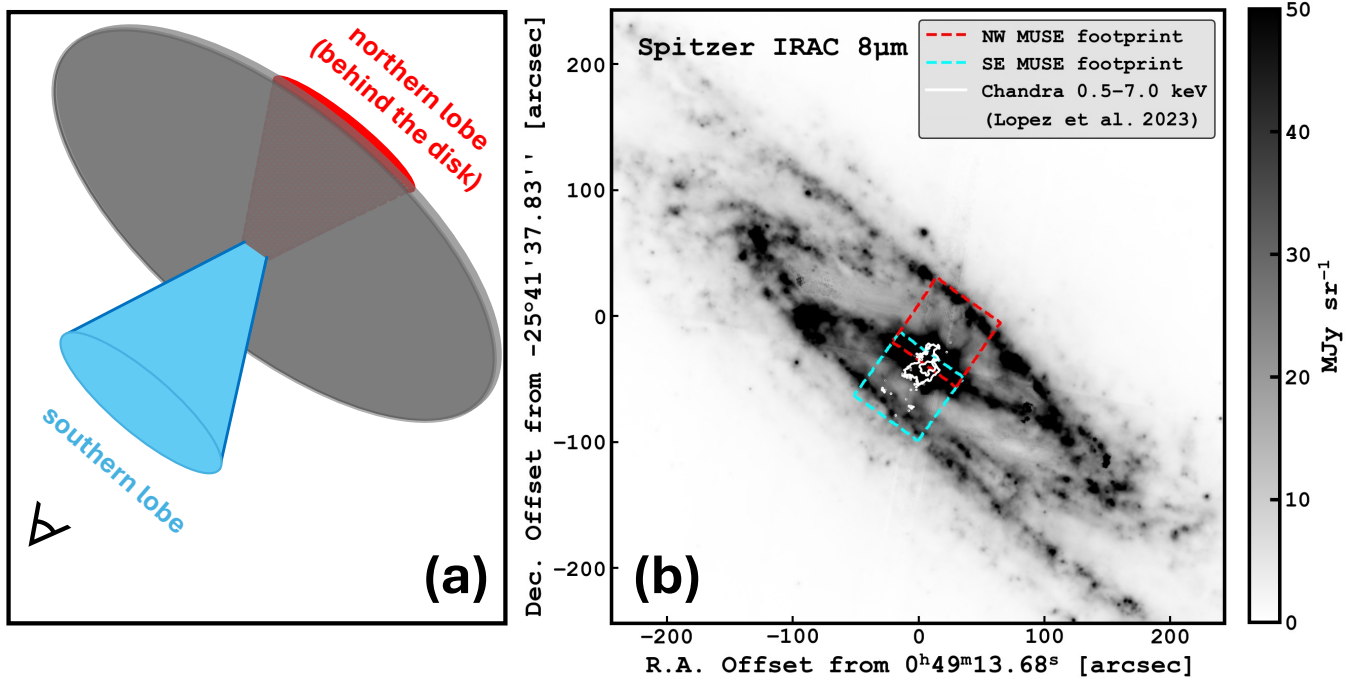

**Figure 1.** (a) A cartoon depicting the geometry of the outflow with respect to the disk in NGC 253. The southern outflow lobe is pointed toward the observer and unobscured by the disk. The northern lobe is receding and behind the disk. (b) Spitzer mid-infrared image of NGC 253 with an overlay of the MUSE footprints. We obtained the IRAC 8 μm data from D. A. Dale et al. (2009) and the NASA/IPAC Extragalactic Database. The footprints of the NW and SE cubes are represented by the red and blue boxes, respectively. The southern portion of the SE cube contains part of the spiral arm, and the entire NW cube is superimposed with the disk and contains the spiral arm at the top. X-ray (0.5–7.0 keV) contours are shown in white (S. Lopez et al. 2023).

In this paper, we take advantage of the spatial and spectral capabilities of the Very Large Telescope's Multi Unit Spectroscopic Explorer (VLT/MUSE) to place detailed constraints on the kinematics and physical properties of the ionized outflow in NGC 253, particularly that of the southern lobe. We first describe the data in Section 2. Then, we jointly model bright Hα, [N II], and [S II] emission lines to decompose outflow and disk components (Section 3). We use these results (Section 4) to determine the physical, deprojected velocities (Section 5.1) and to compare with molecular kinematics (Section 5.2). We then use nebular line diagnostics to place constraints on extinction, the ionization source, and electron densities in the wind (Sections 5.3 and 5.4). We complete our analysis with an estimation of the ionized wind energetics (Section 5.5). A summary of our findings is presented in Section 6. Throughout this work, we assume $V_{sys} = 243$ km s$^{-1}$ (B. S. Koribalski et al. 2004), $d = 3.5$ Mpc (R. Rekola et al. 2005; M. J. B. Newman et al. 2024; S. Okamoto et al. 2024; E. Congiu et al. 2025, in preparation), and galaxy center $\alpha = 00:47:33.1$, $\delta = -25:17:18.6$ (HyperLEDA; D. Makarov et al. 2014).

## 2. Data

MUSE (R. Bacon et al. 2010) is an integral-field spectrograph (IFS) mounted upon the VLT. We obtained two MUSE data cubes as part of program ID 0102.B-0078 (PI: L. Zschaechner). The cubes were observed in the WFM-AO-E mode, which is the Wide-Field Mode with Adaptive Optics (WFM-AO) mode with extended (E) wavelength coverage of 4650−9300 Å. In wide-field mode, MUSE covers a 1 arcmin$^2$ field of view with a continuous spatial sampling of 0″.2, resulting in ∼95,000 spectra per pointing. The two pointings are respectively centered on the outflow lobes in the southeast (SE; $\alpha = 00:47:34.39$, $\delta = -25:17:35.7$) and northwest (NW; $\alpha = 0:47:32.18$, $\delta = -25:16:53.0$). The observations took place on 2018 November 11 (SE) and 2019 July 29 (NW). Each position was observed with a single OB containing four object (O) exposures and two sky (S) exposures, organized in the OSOOSO pattern typical of MUSE observations, for a total of ∼30 minutes on target per field. The observing strategy involved a small dithering pattern and a 90° rotation between exposures to minimize artifacts from the slicers. Each cube has a spectral resolution of $R = 2989$ ($\approx$2.2 Å, $\approx$100 km s$^{-1}$ at Hα). E. Congiu et al. (2025, in preparation) exploited the planetary nebulae they identified in these two cubes to estimate an average point-spread function (PSF) full width at half-maximum (FWHM) at 5000 Å of 0″.7 (NW) and 0″.72 (SE). The raw data were processed by ESO using the standard MUSE pipeline (P. M. Weilbacher et al. 2020).

We overlay the MUSE footprints on top of a Spitzer mid-infrared image to showcase where in NGC 253 the MUSE observations are focused (Figure 1(b)). White contours reveal the extent of 0.5–0.7 keV X-ray emission (S. Lopez et al. 2023) for comparison.

### 2.1. Continuum Subtraction

Removing the underlying stellar continuum is important in deriving line intensities (e.g., E. Bellocchi et al. 2019). To subtract the stellar continuum, we use the PHANGS-MUSE data analysis pipeline, which is described in E. Emsellem et al. (2022). In summary, the pipeline first runs a PPXF-based (M. Cappellari 2017) fitting routine on a Voronoi-binned version of the cube with a reduced number of single stellar population templates recovered from the E-MILES library (A. Vazdekis et al. 2016) to obtain a first guess of the kinematics of the stellar populations. Then, it performs a second fit on the Voronoi-binned data with a more extended sample of templates to extract other physical information (like age and metallicity) of the stellar populations. During this run, the kinematics are fixed to those recovered in the first run. Finally, a third fit is performed, this time on the unbinned data. The first sample of stellar templates is used, and the kinematics remain fixed to those identified in the first run of the fitting routine. The final product of this run is a continuum-subtracted cube, which is what we use for the following analysis.

### 2.2. Ancillary Data

#### 2.2.1. ALMA CO(2–1) Tracing the Molecular Outflow

We aim to compare the kinematics of the warm ionized phase of the outflow with the kinematics of the cold molecular phase as traced by CO(2–1). The CO(2–1) data used in this work were taken during the Atacama Large Millimeter/submillimeter Array (ALMA) Cycle 2 (PI: A. Bolatto, project ID 2013.1.00191.S) and presented in L. K. Zschaechner et al. (2018). The final cube has a spatial resolution of $1\rlap{.}''9 \times 1\rlap{.}''4$ and $5\,\mathrm{km\,s^{-1}}$ spectral resolution. The CO(2–1) emission includes the southwest streamer (SWS) and southeast streamer (SES) that are constrained to the edges of the outflow cones and are also detected in CO(1–0) (A. D. Bolatto et al. 2013) and CO(3–2) (N. Krieger et al. 2019).

#### 2.2.2. JWST/MIRI F770W

We also compare the dust attenuation inferred from our optical measurements with the emission from polycyclic aromatic hydrocarbon molecules at 7.7 μm. We present preliminary $0\rlap{.}''1$ resolution JWST/MIRI 7.7 μm emission in NGC 253 from Cycle 1 GO project 1701 (PI: A. Bolatto).[10] The data were taken with a 4 pt extended dither using the F770W filter. The final image is a 1 × 4 mosaic of the FULL MIRI imaging detector plus a 2 × 2 mosaic of the SUB128 subarray to avoid saturation in the center.

### 3. Method

To determine the properties of the NGC 253 wind, we must fit each spectrum with multiple Gaussian components. A Gaussian distribution is a good model for the underlying instrumental broadening (i.e., the instrumental line-spread function) for MUSE (e.g., R. Bacon et al. 2017). Many spectral lines show velocity contributions from both the outflow and the rotating disk; in some cases, lines of sight may even cut through the front and back walls of the outflow cone, resulting in up to three velocity components per emission line. Thus, we aim to fit each spectrum with one, two, and three "systems of lines." The components of each system must be fit simultaneously, which also reduces the number of degrees of freedom and ensures a fixed separation between the emission lines.

Our routine relies on the spectral cube handling and fitting capabilities of pyspeckit[11] (A. Ginsburg & J. Mirocha 2011; A. Ginsburg et al. 2022) and spectral-cube[12] (T. Robitaille et al. 2016; A. Ginsburg et al. 2019). For this study, we focus on the brightest emission lines: Hα (6562.801 Å), [N II] (6548.05, 6583.45 Å), and [S II] (6716.44, 6730.82 Å). The Gaussian fitter requires initial guesses for each velocity component. Per spaxel, we estimate the disk velocity in two ways: (i) by assuming that the brightest Hα component emits from the disk and not the outflow, and (ii) by adopting a CO(1–0) velocity model of the disk from N. Krieger et al. (2019). The initial guess for the outflow centroid is set to $\pm 50\,\mathrm{km\,s^{-1}}$ relative to this disk velocity. A guess of $-50\,\mathrm{km\,s^{-1}}$ assumes the outflow component arises from the approaching southern lobe, while $+50\,\mathrm{km\,s^{-1}}$ assumes the outflow component comes from either (i) the back wall of the southern cone or (ii) the receding northern lobe. Varying these outflow guesses by 2–4× does not produce different results.

We estimate the initial guesses for the widths ($\sigma$) and amplitudes by the following. We convert the initial guesses for the velocity centroids to wavelength space and calculate $\sigma = \lambda/R/2.355$, where $R \approx 3000$. We remove some degrees of freedom by setting the limit $\sigma_{\mathrm{H}\alpha} = \sigma_{\mathrm{[N\,II]}} = \sigma_{\mathrm{[S\,II]}}$ for each system of lines. We further reduce the number of degrees of freedom by imposing a 3:1 intensity ratio between [N II]$\lambda$6583 and [N II]$\lambda$6548 (e.g., A. Acker et al. 1989). Because the [S II]$\lambda$6716/[S II]$\lambda$6730 line ratio is a diagnostic of electron density ($n_e$), these amplitudes are instead allowed to float in parameter space. We determine the initial guesses for amplitude by visual inspection, setting each component of ([N II]$\lambda$6548, Hα, [N II]$\lambda$6583, [S II]$\lambda$6716, [S II]$\lambda$6730) = (100, 300, 300, 150, 150 $10^{-20}\,\mathrm{erg\,s^{-1}\,cm^{-2}\,\mathring{A}^{-1}}$), and find that varying these inputs does not change our results.

Figures 2 and 3 display a series of example spectra (in black) and their favorable fits (components in blue, composite in pink) in both cubes. We also display the residuals in purple. The spectra are chosen to showcase the complexity of the emission lines: high and low S/N, symmetric and asymmetric shapes, and single- and multi-peaked profiles.

### 3.1. Decomposing Emission Lines

We now must determine which fit (one, two, or three systems of lines) best models each spectrum. For each fit, we calculate the chi-square ($\chi^2$) and obtain the Bayesian information criterion (BIC):

$$\mathrm{BIC} = \chi^2 + n \times \ln(m), \quad (1)$$

where $n$ is the number of free parameters (6, 12, and 18 for one, two, and three systems of lines, respectively) and $m$ is the number of channels ($m = 150$). The BIC is a similar diagnostic to the reduced chi-square $\chi_r^2$ ($\chi^2$ divided by the number of degrees of freedom), but also penalizes models that introduce more parameters. Generally, the smaller the BIC, the better the fit.

We use the BIC to determine on a spectrum-by-spectrum case the number of components needed to decompose the emission lines. We first assume that the one-system-of-lines model is best everywhere and only prefer the two-system model when its BIC value, $BIC_2$, is less than $BIC_1$. A three-system model is then preferred when $BIC_2/BIC_3 > 1.5$. We experiment with different thresholds and visually inspect the resulting maps. We find that generally $BIC_1/BIC_2 > 1$ works well when preferring two components. We take a more conservative approach when determining between two and three components, because a $BIC_2/BIC_3$ threshold less than 1.5 tends to result in overfitting, whereas a threshold greater than 1.5 often underfits spectra with obvious outflow components. Being more strict when adopting three components is consistent with M. S. Westmoquette et al. (2011), who found that most lines can be fitted with one or two components.

---

[10] The JWST data used in this paper can be found in MAST: 10.17909/qpdy-1n82.

[11] https://pyspeckit.readthedocs.io/en/latest/
[12] https://github.com/radio-astro-tools/spectral-cube

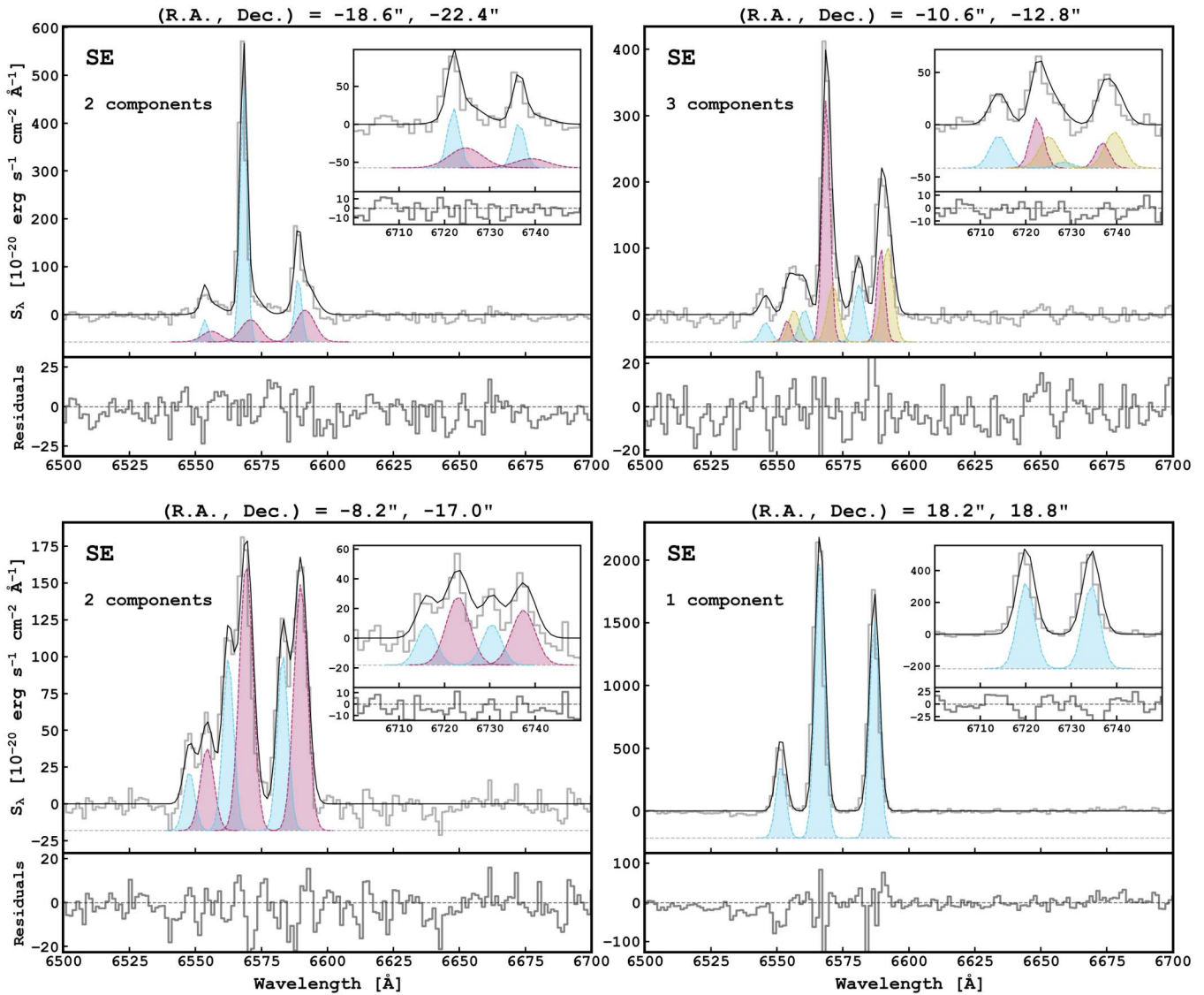

**Figure 2.** Example of fitted spectra in the SE data cube, labeled by R.A. and decl., and showing the preferred number of components based on Figure 4. The main plot features the Hα−[N II] complex, and the inset is a zoom in on the [S II] doublet with the same R.A. and decl. offset labels. The individual components (colored dashed curves), which are offset by 10% of the brightest line, add up to the composite fit (black curve). We color the components to make it easier to discern each system of lines. The residuals can be found below each spectrum in gray. We choose these particular spectra to showcase the complexity and variety of the data.

The BIC threshold does not always work. In spectra with high signal-to-noise (S/N), the BIC calculation sometimes favors ⩾2 components when visually it is not apparent that ⩾2 components are needed. This becomes obvious when the velocity separation between the components is small. Thus, if the velocity separation between multiple components is less than a channel width (i.e., $<57\,km\,s^{-1}$), we determine that the spectrum is better fit with the one-system-of-lines model.

Figure 4 illustrates the number of components per spaxel based on the BIC and velocity separation criteria. The majority of spectra in the NW cube can be modeled with one system of lines. There are patches of spaxels that require two- and three-line systems, but it is not immediately evident that these regions are truly associated with the northern outflow lobe. In the SE cube, the shape of the outflow cone is already apparent, with the majority of the cone favoring two components. Three components are preferred in patches toward the bottom of the map. We also find some spaxels best fit with three components toward the nucleus of NGC 253, which we suspect may be due to overfitting high-S/N regions that were not caught by either the BIC or velocity separation tests.

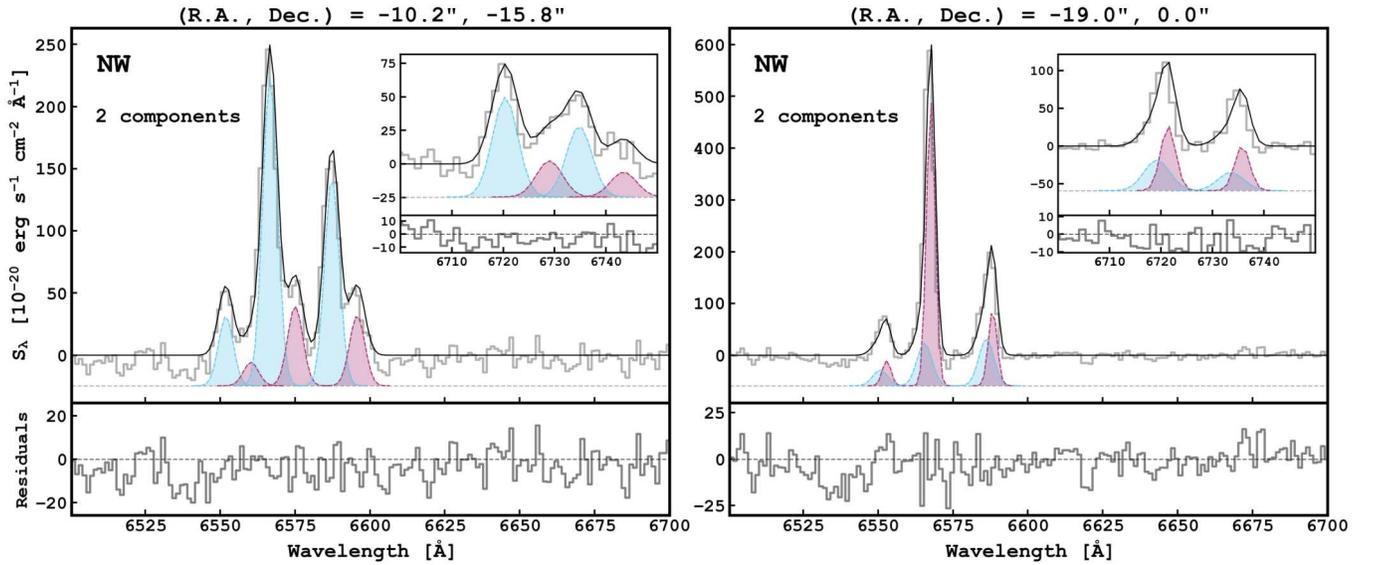

**Figure 3.** Same as Figure 2, but for two example NW spectra with possible outflow features. The vast majority of NW spaxels are only fit with a single disk component (not shown).

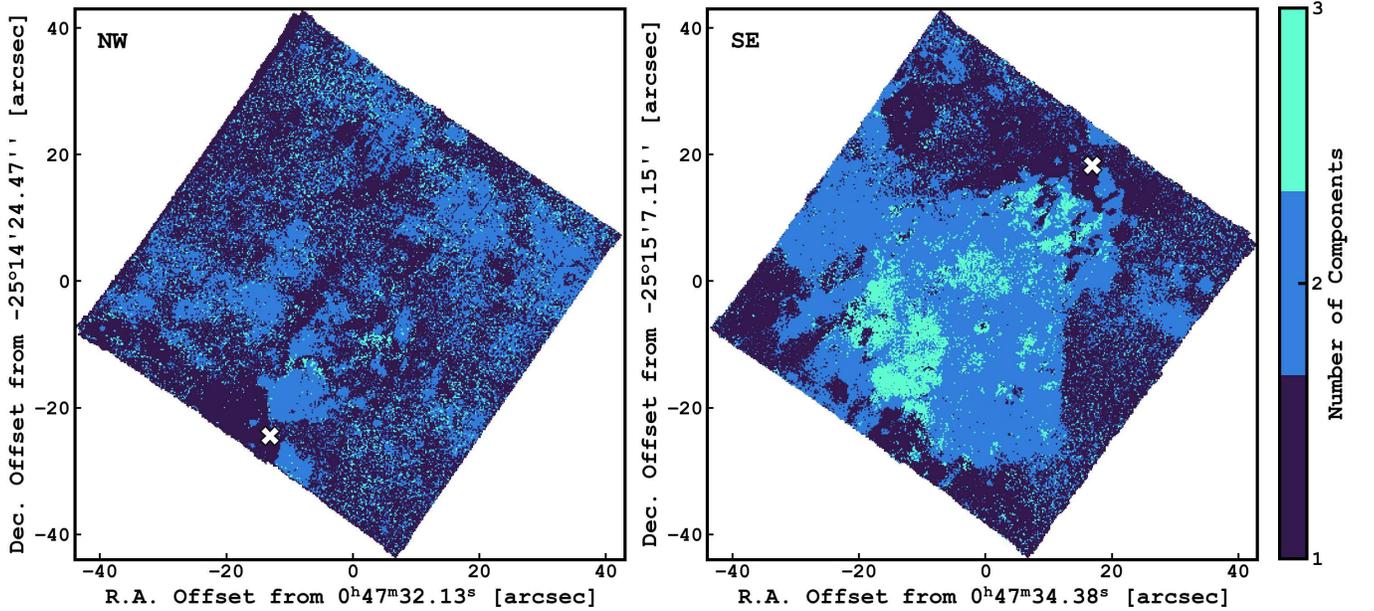

**Figure 4.** The number of components (or the number of systems of lines) that best fit the spectrum in each spaxel (see Sections 3.1 and 3.2 for details.) The white × marks the galaxy center. The shape of the southern (SE) outflow cone is already apparent when considering which spaxels require ⩾2 components.

### 3.2. Separating Components

After determining the number of components that best models each spectrum, we must deduce which components are due to the outflow and which are caused by the rotation of the disk. For those spectra best modeled with one system of lines, the sole velocity component is assigned to the disk. For those with >2 components, we use the assumption that the brightest Hα component arises from the disk. We also cross-check this with the CO(1−0) velocity map from N. Krieger et al. (2019). In the two-systems-of-lines fit, the second component is assigned to the "Blue Outflow" map. In the case of the three-systems-of-lines fit, we split up the outflow components based on which is more blueshifted ("Blue Outflow") and which is more redshifted ("Red Outflow") with respect to the disk component.

## 4. Results

### 4.1. Intensities

Figure 5 displays the total continuum-subtracted flux summed over each emission line in both cubes. The NW data set is brightest in Hα due to the intervening disk. Knots of ionized emission are most notable in the northern spiral arm. In the SE data, similar bright knots match up with known H II regions measured by K. Matsubayashi et al. (2009) using the Subaru Telescope. This same study also found a filament ("filament C") that is distinct from the outflow, which appears bright in our data. The large filamentary structures extending from the nucleus to the southern spiral arm match up with the limb-brightened outflow cone (T. M. Heckman et al. 1990), and are precisely the locations of the known molecular SWS and SES (A. D. Bolatto et al. 2013; F. Walter et al. 2017). The outflow cone appears brightest in [N II].

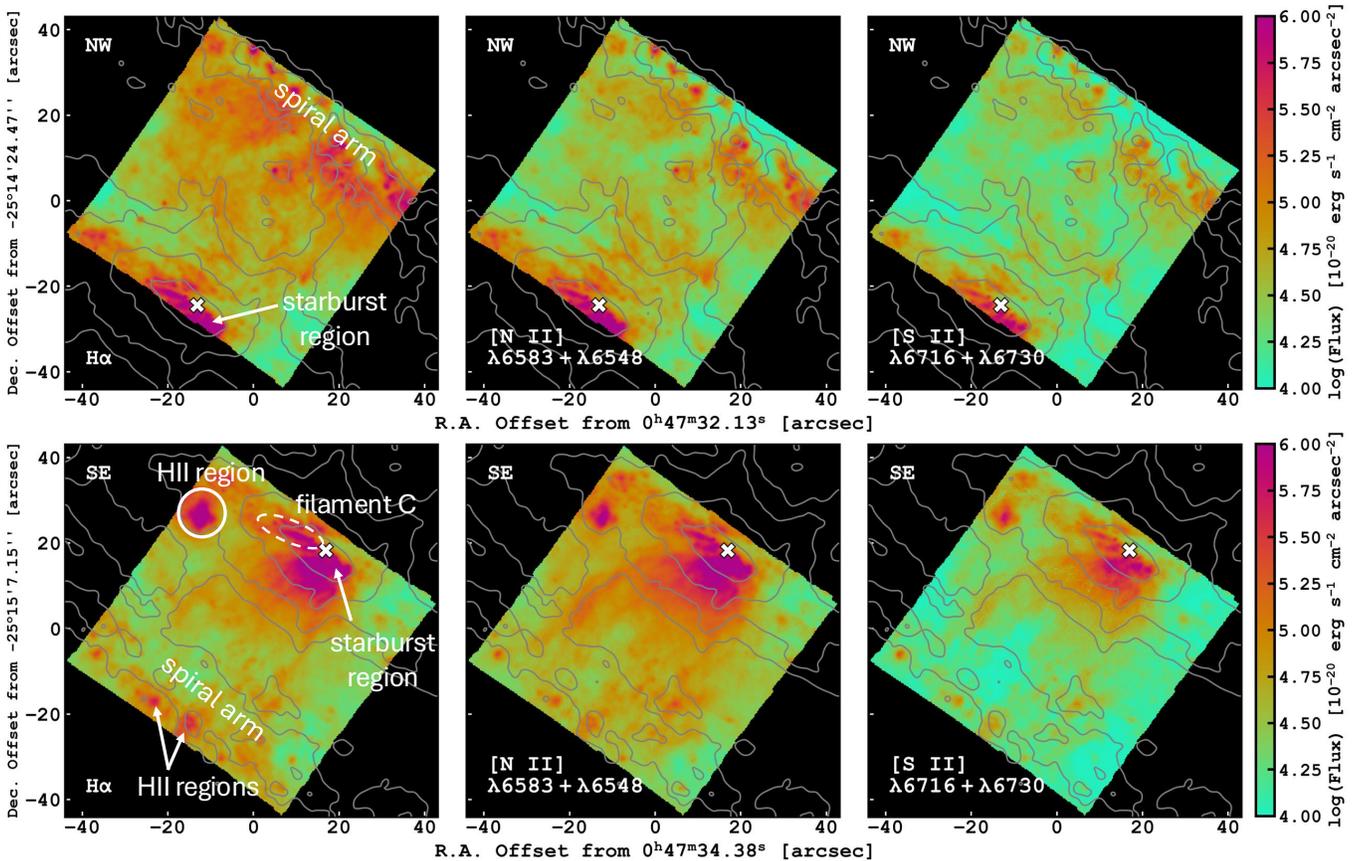

**Figure 5.** Total flux of each emission line in both the NW (top panels) and SE (bottom panels) MUSE cubes. Both cubes overlap at the nucleus (the white × indicates the galaxy center). White ovals and arrows mark known disk features from K. Matsubayashi et al. (2009), including H II regions, the starburst, and "filament C," which they find to be kinematically distinct from the outflow. Outflow features are already apparent in the SE data. Gray contours from JWST/MIRI 7.7 μm emission (levels = [30, 50, 155, 1023] MJy sr$^{-1}$) trace the nucleus, bar, and outflow filaments that stretch to the spiral arms.

Figure 6 presents the integrated flux for each component in the SE cube. The H II regions and "filament C" from K. Matsubayashi et al. (2009) are successfully assigned to the disk rather than the outflow, confirming our method of component separation from Sections 3.1 and 3.2. The Blue Outflow map contains the majority of outflow components, with an obvious cone shape. Here we find the wind to be brightest in Hα and [N II]. Much of the emission is nonuniform throughout the Blue Outflow maps, indicating that the ionized cone (especially the Hα-emitting gas) is clumpy.

The components in the Red Outflow are mainly found in a patch near the southern spiral arm. This suggests that these components arise from confusion with the arm instead of the actual outflow. It is possible, however, that we are seeing the back wall of the outflow cone at this location.

### 4.2. Projected Velocities

A galactic wind can be modeled as a truncated cone with the frustum located at the wind-launching site. A schematic of an outflow cone is illustrated in Figure 7, modeled as a full, nontruncated cone for clarity on the angles. The outflow cone has a semi-opening angle $\alpha$, which determines how much the front and back walls of the cone are projected toward and away from the observer supplementary to any non-face-on geometry of the galaxy. In the case of NGC 253, kinematic modeling has revealed a semi-opening angle (gray arc) $\alpha = 30°$ and inclination (purple arc) $i \approx 78°$ (M. S. Westmoquette et al. 2011). We discuss this geometry and its effects on the observed kinematics in detail in Section 5.1.

Figure 8 features the projected velocities ($v_{\mathrm{proj}}$) and intrinsic FWHM measurements of the NGC 253 outflow and disk components in the SE. The velocities are projected, i.e., not yet corrected for the inclination of both the disk and outflow cone (see Section 5.1). We choose to display the [N II]λ6583 results as we expect this line to be bright in the outflow. [N II]λ6548, Hα, and the [S II] doublet exhibit the same kinematics as they are all part of the same system of gas.

The projected velocities vary smoothly across the outflow and disk maps. We find that the southern lobe of the outflow is best represented by the Blue Outflow map. The general shape of the Blue Outflow map reflects that of a conical frustum, and here the outflowing gas is largely blueshifted with respect to the disk. This result is expected given that the southern outflow cone is pointed toward the observer (see the geometry in Figure 1). Furthermore, the edges of the Blue Outflow cone match up with the known locations of the SES and SWS (A. D. Bolatto et al. 2013; F. Walter et al. 2017). These streamers are highlighted by the JWST/MIRI 7.7 μm emission contours overlaid in Figure 8. We consider the area between the SES and SWS of the Blue Outflow map to be the southern outflow cone. The material outside this area may be due to motions along the bar and confusion with the southern spiral arm. This is also the case with the Red Outflow components, which are sparse and superimposed on the southern spiral arm.

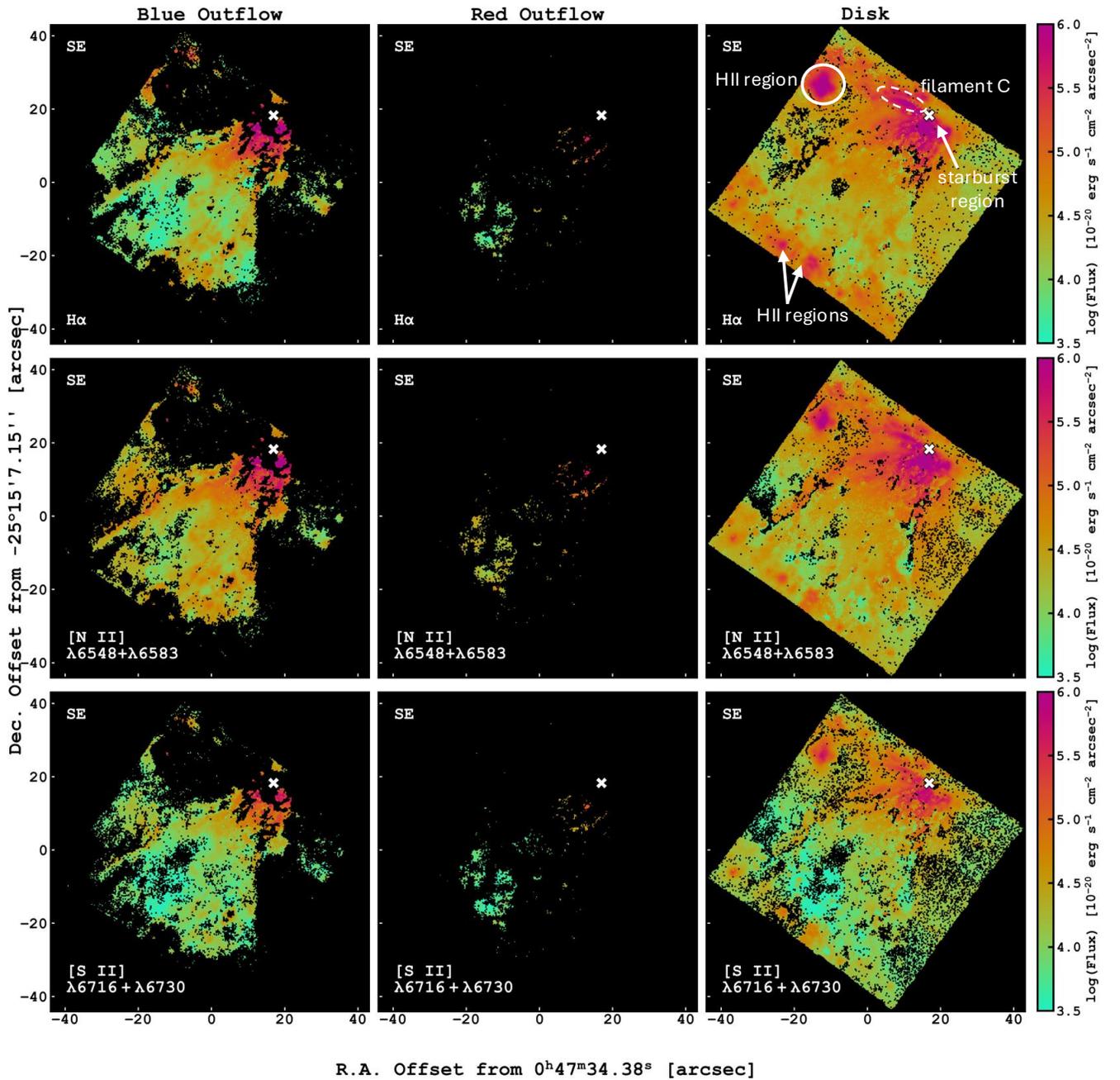

**Figure 6.** Integrated flux of Hα, [N II]λ6548+[N II]λ6583, and [S II]λ6716+[S II]λ6730 in each component in the SE cube. The white × marks the galaxy center. All emission lines are brightest in the H II regions and the starbursting nucleus. The emission is not uniformly bright throughout the Blue Outflow, indicating that the outflow cone is clumpy.

In the southern outflow cone, the average velocity is $\sim-150$ to $-250\,\mathrm{km\,s^{-1}}$ with respect to systemic. In the bluest part of the southern outflow lobe, where the cone may be pointed most toward the observer, the average velocity is $\sim-275\,\mathrm{km\,s^{-1}}$ with respect to systemic.

Similar to Figure 8, Figure 9 presents the ionized gas kinematics targeting the northern lobe of the outflow. We find very little spaxels that require three components, and thus only present one outflow map. The northern outflow lobe is detected through the disk in two patchy regions. The redshifted clump of emission at $(x, y) \approx (-10'', -20'')$ contains clear double-peaked emission lines (see the left panel of Figure 3). Here the outflow velocities range from $\sim100$ to $300\,\mathrm{km\,s^{-1}}$. The blueshifted clump around $(x, y) \approx (-20'', 0'')$ contains asymmetric line profiles (see the right panel of Figure 3) with outflow velocities $\sim-150\,\mathrm{km\,s^{-1}}$. This clump falls within the contours of a northern streamer detected at 7.7 μm with JWST/MIRI and also in CO (A. D. Bolatto et al. 2013; F. Walter et al. 2017). Where the NW and SE pointings overlap in the nucleus, the disk kinematics are consistent, with an average $v_{\mathrm{NW,disk}}/v_{\mathrm{SE,disk}} = 0.3\%$.

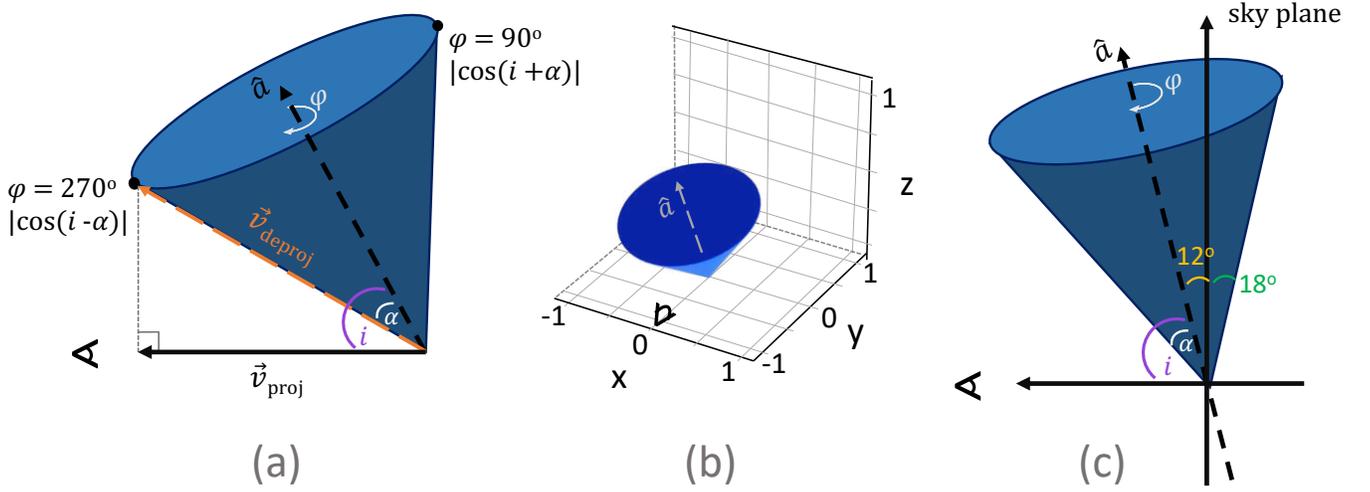

**Figure 7.** Geometry of a cone with inclination $i$ and semi-opening angle $\alpha$ measured with respect to its rotational axis ($\hat{a}$, dashed line). (a) As a function of $\varphi$ (the azimuthal angle around the cone axis), $v_{\mathrm{deproj}}$ traces the wall of the cone, and the projected velocity vector $v_{\mathrm{proj}}$ (solid black line) lies in the $X$–$Y$ plane. Here $v_{\mathrm{deproj}}$ points toward the observer at $\varphi = 270°$ (left black point), in which the velocity has a maximum projection factor ($|\cos(i - \alpha)|$) toward the observer. The maximum projection factor away from the observer is then $|\cos(i + \alpha)|$ at $\varphi = 90°$ (right black point). (b) A 3D cone in the $X$–$Y$–$Z$ plane with the observer at $X = 0$. (c) The same cone as (a), but noting the angle projected from the sky plane. For $i = 78°$, this angle is small (12°), which is why most lines of sight do not detect the back wall of the cone.

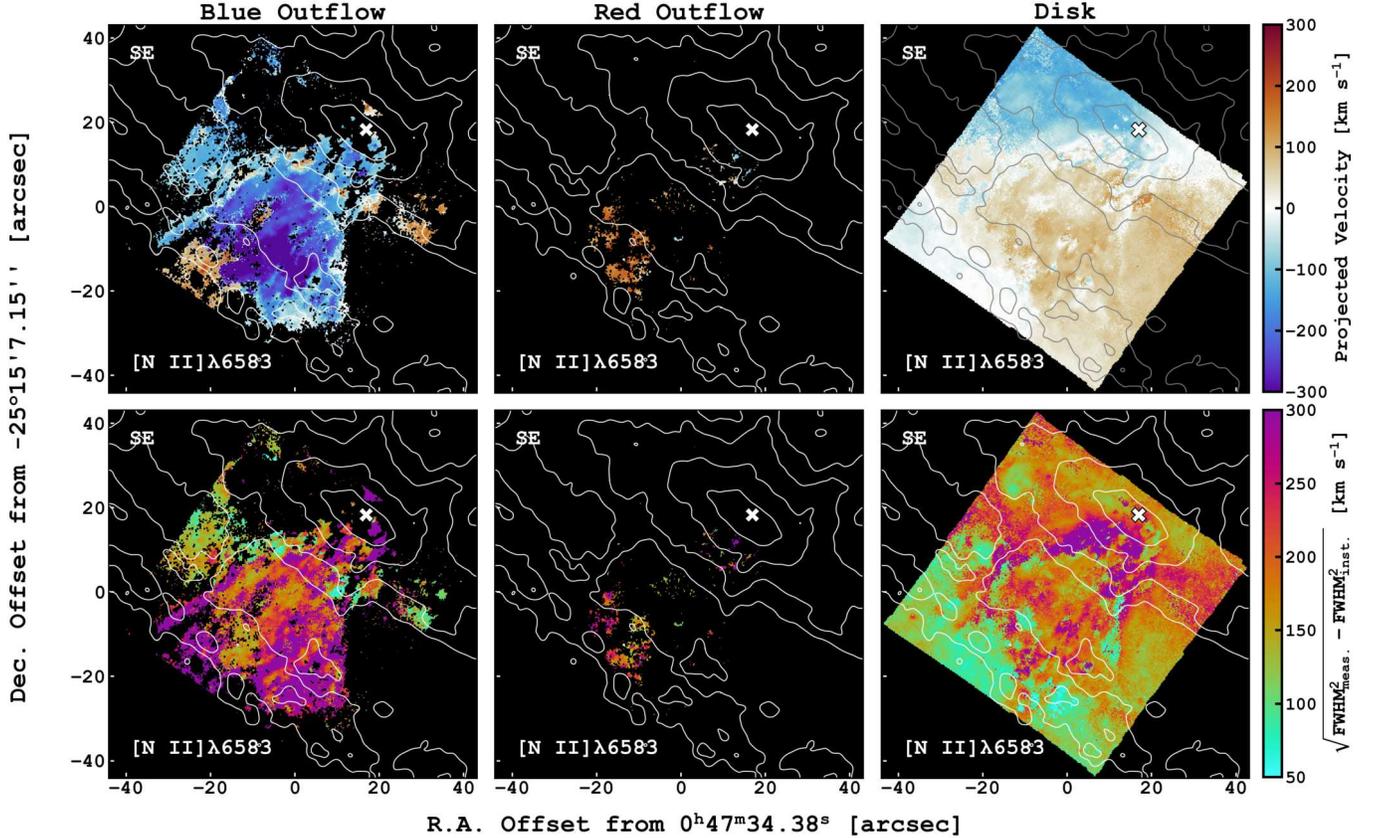

**Figure 8.** Kinematics of the southern outflow lobe in NGC 253. Top: projected [N II]$\lambda 6583$ velocities of the main outflow cone ("Blue Outflow"), additional redshifted outflow components ("Redshifted Outflow"), and disk components ("Disk"). The average projected velocities of the southern outflow cone are between $\sim -150$ and $-250$ km s$^{-1}$. In the bluest region of the Blue Outflow map, velocities reach $\sim -350$ km s$^{-1}$. Redshifted components in both outflow maps may be caused by confusion with the intervening disk. Bottom: intrinsic FWHM calculated by subtracting in quadrature the instrumental resolution (100 km s$^{-1}$) from the measured FWHM. The average Blue Outflow line-width map is $\sim 250$ km s$^{-1}$. Broad line widths may indicate that the ionized gas is filling the outflow cone. Both: JWST/MIRI F770W contours highlight the nucleus, spiral arms, and outflow filaments. The white × marks the galaxy center.

### 4.3. Line Widths

The southern outflow in NGC 253 is known to contain large line widths that are speculated to be caused by gas filling the outflow cone (M. S. Westmoquette et al. 2011). Intrinsic FWHM measurements in the southern outflow are found in the bottom panels of Figure 8. We calculate the intrinsic FWHM by subtracting in quadrature the instrumental resolution ($\approx 100$ km s$^{-1}$) from the fitted FWHM. The mean Blue Outflow and Disk intrinsic FWHMs are $\approx 250$ and 170 km s$^{-1}$, respectively. The broadest FWHM in the outflow cone is $\sim 500$ km s$^{-1}$, consistent with the $>400$ km s$^{-1}$ H$\alpha$ line widths measured by M. S. Westmoquette et al. (2011). In the northern lobe, the red clump of outflow material also reaches line widths $\sim 500$ km s$^{-1}$. The average broadening in the blueshifted clump is between $\sim 200$ and 250 km s$^{-1}$. These broad line widths are further evidence that some lines of sight are cutting through patchy regions in the disk and detecting parts of the northern outflow lobe, which to our knowledge has not been seen before at optical wavelengths in NGC 253.

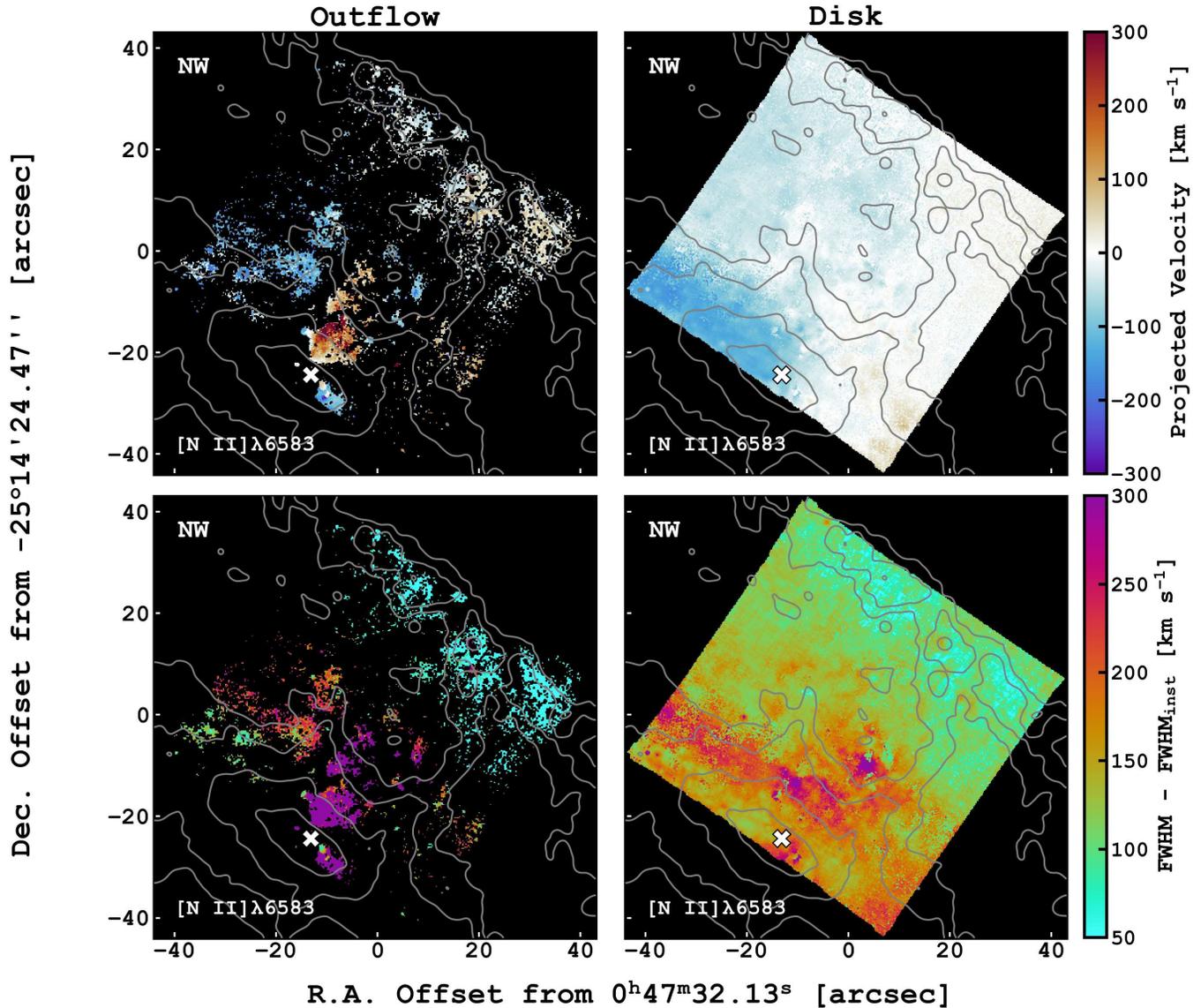

**Figure 9.** Same as Figure 8, but for the NW data set. We exclude the Red Outflow map because only a sparse number of spaxels could be fit with three components. Here we overlay JWST/MIRI F770W contours (levels = [30, 50, 155, 1023] MJy sr$^{-1}$). We detect patches of the northern outflow lobe in the redshifted clump near $(x, y) \approx (-10'', -20'')$ and the blueshifted material near $(x, y) \approx (-20'', 0'')$. Outflow velocities range from $\sim -150$ km s$^{-1}$ to $+300$ km s$^{-1}$. Line broadening reaches $\sim 500$ km s$^{-1}$ in the redshifted patch of the northern outflow, suggesting that this material may be filling the cone.

## 5. Discussion

Our analysis focuses on the well-detected outflow components in the SE. Specifically, we refer to the SE Blue Outflow maps when regarding the southern outflow lobe.

### 5.1. Deprojected Velocities

The geometry of the southern outflow cone causes projection effects in the kinematics (Figure 7). Therefore, the outflow velocities discussed in Section 4.2 are lower limits on the true wind speeds. In this section, we use the discussion in A. D. Bolatto et al. (2021) to calculate the intrinsic, "deprojected" velocities ($v_{\mathrm{deproj}}$) along the SE outflow cone. We ignore the thickness of the ionized gas.

Panel (a) in Figure 7 illustrates the geometry of an inclined outflow cone with a semi-opening angle ($\alpha$) that is measured from the cone's major axis ($\hat{a}$). The vector $v_{\rm deproj}$ is drawn so that it traces the outer wall of the cone as a function of $\varphi$: the azimuthal angle around the cone's major axis. If the distribution of material in $\varphi$ is uniform, an observer—who sees a 2D projection of the 3D cone (i.e., in the $X$–$Z$ plane in Figure 7(b))—will see an excess of emission at $\varphi = 0°$, 180°, or the cone edges. This geometry leads to the limb-brightening phenomenon seen in both the molecular and ionized phases (see discussion in A. D. Bolatto et al. 2021).

$v_{\rm proj}$ is the projection of $v_{\rm deproj}$ onto the $X$–$Y$ plane, specifically in the $-y$ direction toward the observer (see Figure 7(b)): $v_{\rm proj} = v_{\rm deproj} \cdot -\hat{y} = |v_{\rm deproj}| \cos\theta$. Here $\theta$ takes into consideration both $i$ and $\alpha$. Figure 7(a) portrays the simplest case in which $\vec{v}_{\rm deproj}$ is pointed most toward the observer. This occurs at $\varphi = 270°$, and means $\cos\theta = \cos(i - \alpha)$. For the part of the cone pointed most away from the observer, $\varphi = 90°$ and $\cos\theta = \cos(i + \alpha)$. With $i = 78°$ and $\alpha = 30°$, the projection factor at $\varphi = 270°$ is $|\cos(i - a)| = 0.67$; for $\varphi = 90°$, the projection factor is $|\cos(i + a)| = 0.31$.

At the center of the Blue Outflow map, where we expect the cone to be most pointed toward us ($\varphi = 270°$), projected velocities reach average speeds of $v_{\rm proj} \approx 275 \pm 70 \,\rm km\,s^{-1}$. This equals average deprojected velocities of $v_{\rm deproj} \sim -400 \,\rm km\,s^{-1}$. Some of the highest projected velocities in this region reach $v_{\rm proj} \sim -350 \,\rm km\,s^{-1}$, or $v_{\rm deproj} \sim -500 \,\rm km\,s^{-1}$.

In the galactic fountain model, gas that does not exceed the escape velocity of the galaxy will accrete back onto the disk (T. M. Heckman et al. 1990; S. Veilleux et al. 2020). In NGC 253, the hottest ($\sim 10^7$ K) gas most likely can reach these speeds, as X-ray lobes have been detected out to $\sim 10$ kpc (D. K. Strickland et al. 2000, 2002; I. Mitsuishi et al. 2013). The escape velocity of NGC 253 is estimated to be from a few hundred kilometers per second up to $\sim 500 \,\rm km\,s^{-1}$ (e.g., V. Heesen et al. 2009; F. Walter et al. 2017). Adopting this estimate, it is apparent that a considerable amount of the outflowing H$\alpha$-emitting material could escape NGC 253, especially if the material is indeed accelerating with distance from the midplane, which we explore in Section 5.2.

The projection of the cone in the plane of the observer may answer a question about component modeling and separation: Why do most lines of sight not detect the back wall of the outflow cone? Without taking geometry into consideration, it is not obvious why there would only be a velocity contribution from the front wall. Panel (c) in Figure 7 illustrates the differences in projection effects from the back wall of the cone versus the front. Measuring the inclination from the rotation axis $\hat{a}$, the angle between $\hat{a}$ and the plane of the sky is $90° - i = 90° - 78° = 12°$. This means that the back wall of the cone (which is behind the plane of the sky) is only projected at $\alpha - 12° = 30° - 12° = 18°$ toward the observer. Compare this projection to the total $12° + 30° = 42°$ pointed toward the observer from the front wall of the cone, and we can see that the back wall of the cone only produces a small contribution to the overall projected outflow velocity, making it difficult to disentangle from the disk. Therefore, it is expected that the vast majority of the velocity contribution comes from the front wall of the outflow cone.

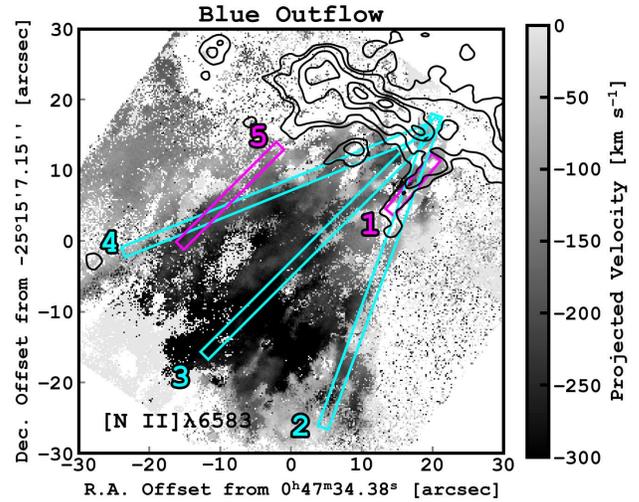

**Figure 10.** Slices along the SE Blue Outflow map. The slices are numbered 1–5, with slices 1 and 5 (magenta) corresponding to locations near the molecular southwest streamer (SWS) and southeast streamer (SES), respectively (L. K. Zschaechner et al. 2018; see CO(2–1) contours). Slice 1 has a length of $10\farcs2$, and slice 5 has a length of $20''$. Slices 2–4 (cyan) extend radially from the base of the outflow cone and each span $47''$ in length. Each slice is $1\farcs54$ wide ($7.7\times$ the pixel scale).

### 5.2. Evidence for Acceleration and Comparison with Molecular Kinematics

In a galactic wind, the "cloud crushing" timescale of cold, dense molecular clouds is expected to be less than the time it takes to be accelerated by a hot wind ($t_{\rm cc} < t_{\rm accel}$). How, then, does molecular gas ever survive long enough to escape a galaxy? One possible solution, as brought forth by M. Gronke & S. P. Oh (2018, 2020), is that the time it takes for the hot wind to cool and mix with the cold gas ($t_{\rm cool,mix}$) is less than $t_{\rm cc}$. In this case, the hot gas mixes with the cold gas, cools, and replenishes the cold reservoir. The hot wind carries enough momentum to then entrain the mixed cold reservoir and accelerate it out of the galaxy.

To explore this idea, we seek to compare velocity gradients of both the ionized and molecular phases of the NGC 253 wind. These velocity gradients could provide insight into the acceleration of cold gas by a hot wind. In Figure 10, we overlay CO(2–1) contours from L. K. Zschaechner et al. (2018) onto a grayscale version of our Blue Outflow map. Here we see the overlap between the ionized gas and the molecular SWS; unfortunately, we do not have optical outflow measurements along the area associated with the molecular SES. Note that the ionized gas measurements along the SWS are not continuous but instead patchy.

We take a series of $1\farcs5$-wide slices (about 8 pixels wide) across the Blue Outflow map (magenta and cyan rectangles in Figure 10). Slice 1 (magenta) corresponds with the location of the molecular SWS, is $10\farcs2$ in length, and is the only slice with significant overlap with the CO(2–1) observations. Slices 2–4 (cyan) are $47''$ in length and are positioned to capture possible gradients at both the edges and down the center axis of the cone. Slice 5 (magenta) is $20''$ long and is near the location of the molecular SES.

In Figure 11, we plot the $v_{\rm proj}$ of the SE Blue Outflow map as a function of distance along the length of each slice running north to south. These measurements are averaged along the width of the slice and are represented by the light gray data points. To show the global trends, we plot the mean $v_{\rm proj}$ in 15 bins and overlay them in the colored points. In the slices where we observe clear velocity gradients, we plot a best-fit line.

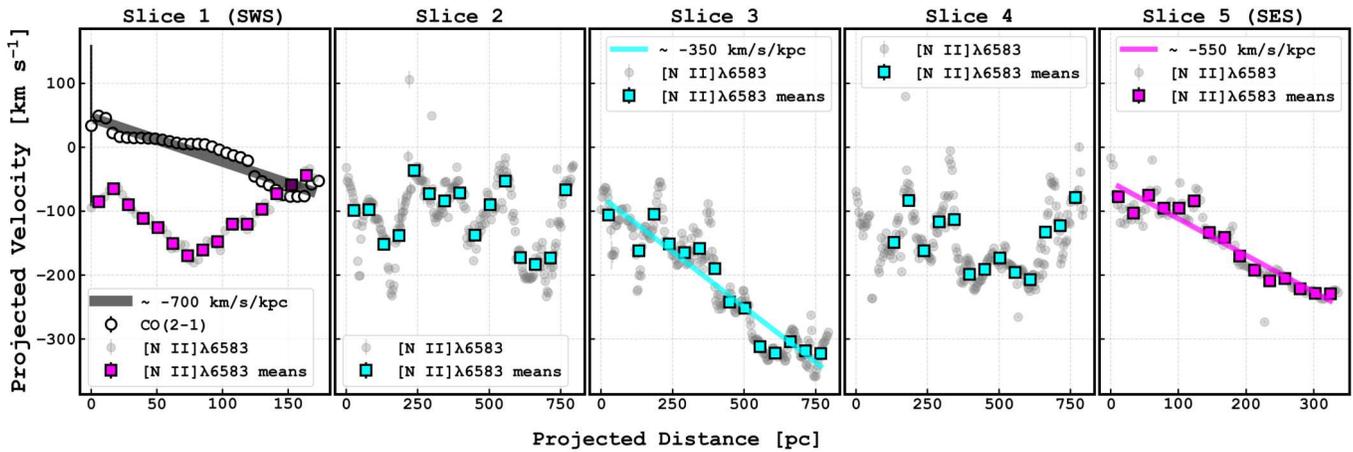

**Figure 11.** [N II]λ6583 projected velocities measured from north (0 pc) to south along the slices in Figure 10 (magenta indicating the SES and SWS). The gray data points are velocities averaged along the width of the slice. We then bin these data and plot the mean of each bin in the colored squares. For the slices that show straightforward velocity gradients, we fit a line to the mean of the bins (thick lines). We find a velocity gradient of $\sim-350$ km s$^{-1}$ kpc$^{-1}$ in slice 3 and $\sim-550$ km s$^{-1}$ kpc$^{-1}$ in slice 5. To compare with molecular gas in slice 1 (the SWS), we calculate a gradient of $\sim-700$ km s$^{-1}$ kpc$^{-1}$ in CO(2–1) with data from L. K. Zschaechner et al. (2018). Here the [N II] data show a reversal in the velocity gradient seen with CO, likely due to extinction. Most error bars are smaller than the points.

We also calculate CO(2–1) $v_{\rm proj}$ along the SWS (slice 1) by first averaging together the spectra along the width, and then performing a two-component Gaussian fit. We assume that the blueshifted component results from the outflow, informed by the findings of L. K. Zschaechner et al. (2018). For this calculation, we use the same $10'' \times 1\rlap{.}''5$ slice as before, where the width roughly matches the beam size of the CO(2–1) cube ($1\rlap{.}''9 \times 1\rlap{.}''4$). These measurements are represented by open circles in the first panel of Figure 11.

First, we look for evidence of acceleration in [N II] along the outflow cone. We fit a clear velocity gradient along slices 3 and 5, both accelerating to $\sim-350$ km s$^{-1}$ kpc$^{-1}$ and $\sim-550$ km s$^{-1}$ kpc$^{-1}$, respectively. This is in agreement with M. S. Westmoquette et al. (2011), who note that ionized gas velocities seem to increase with distance along the outflow cone. We do not find clear gradients in slices 1, 2, or 4.

We next compare the CO(2–1) and ionized gas velocities in the SWS (slice 1). We find a relatively steady velocity gradient ($\sim-700$ km s$^{-1}$ kpc$^{-1}$) in CO(2–1). Though our slice does not exactly match that of F. Walter et al. (2017), $-0.7$ km s$^{-1}$ pc$^{-1}$ is consistent with their estimate of the molecular gas accelerating at $-1$ km s$^{-1}$ pc$^{-1}$. Such a trend in [N II] is much less clear for the SWS. We expect [N II] to be moving faster than CO(2–1), as the ionized gas should entrain and accelerate the molecular gas. Instead, we only capture such a pattern up to $\sim75$ pc from the midplane, where the ionized gas motions then show a partial reversal of the velocity gradient observed in CO. The SWS is a dusty feature, and so we suspect that extinction may be playing a role in this type of behavior.

While our comparison between phases is inconclusive, we do find that some of the ionized gas is accelerating. When comparing to the NGC 253 escape speed of 500 km s$^{-1}$, the acceleration we find is further evidence that at least part of the H$\alpha$-emitting wind is likely to escape the galaxy.

### 5.3. Extinction

In the previous section, we speculate that some of our velocity gradient measurements suffer from nonuniform high extinction. Similarly, we find challenges in the emission-line decomposition that may also be due to high levels of extinction (see Section 4.2).

In the top panels of Figure 12, we plot the integrated flux Balmer decrement H$\alpha$/H$\beta$, where a nominal value of 2.86 (0.44 in log-space) indicates significant reddening. For this calculation, we use the Gaussian decomposition results from H$\alpha$ to inform the number of components needed and the location of the velocity centroid for H$\beta$. We then sum over these components to calculate the total integrated flux. We find that a substantial fraction of both the NW and SE pointings suffer from reddening. Our estimation of extinction assumes screen geometry, which may not be appropriate for the NW pointing because there is mixing in the screen between the outflow and the disk. This may be why, excluding the wind component, the NW values appear smaller than the SE.

In the bottom panels of Figure 12, we convert the H$\alpha$/H$\beta$ measurements to extinction magnitudes $A_V$. Assuming Case B recombination and $T_e = 10^4$ K, the color excess can be measured as

$$E(B - V) = \frac{2.5}{k({\rm H}\beta) - k({\rm H}\alpha)} \log_{10}\left(\frac{F_{{\rm H}\alpha}/F_{{\rm H}\beta}}{2.86}\right). \quad (2)$$

Assuming a D. Calzetti et al. (2000) attenuation law (their Equation (4)), and $R_V = 3.1$ (standard for the diffuse ISM), $k({\rm H}\alpha) = 3.65$ and $k({\rm H}\beta) = 2.38$. From D. E. Osterbrock & G. J. Ferland (2006), $A_V = R_V E(B - V)$, which can be found in the bottom panels of Figure 12.

In the SE, the median $A_V \sim 1$ mag but reaches $A_V < 1$ mag in locations corresponding with the southern outflow cone. This is lower than the $A_V$ found in the MAGNUM survey of outflows in local active galaxies ($A_V \geqslant 2$ mag; M. Mingozzi et al. 2019). In the nucleus of NGC 253, extinction magnitudes exceed $A_V \geqslant 5$. We note that the NW extinction values appear on average lower than the SE. This is again likely due to the assumption of screen geometry in the H$\alpha$/H$\beta$ calculation. The SE extinction map also appears more patchy than the NW. This suggests that the dust is nonuniform in the outflow cone, which could be contributing to the lack of smooth velocity features and gradients explored in Section 5.2.

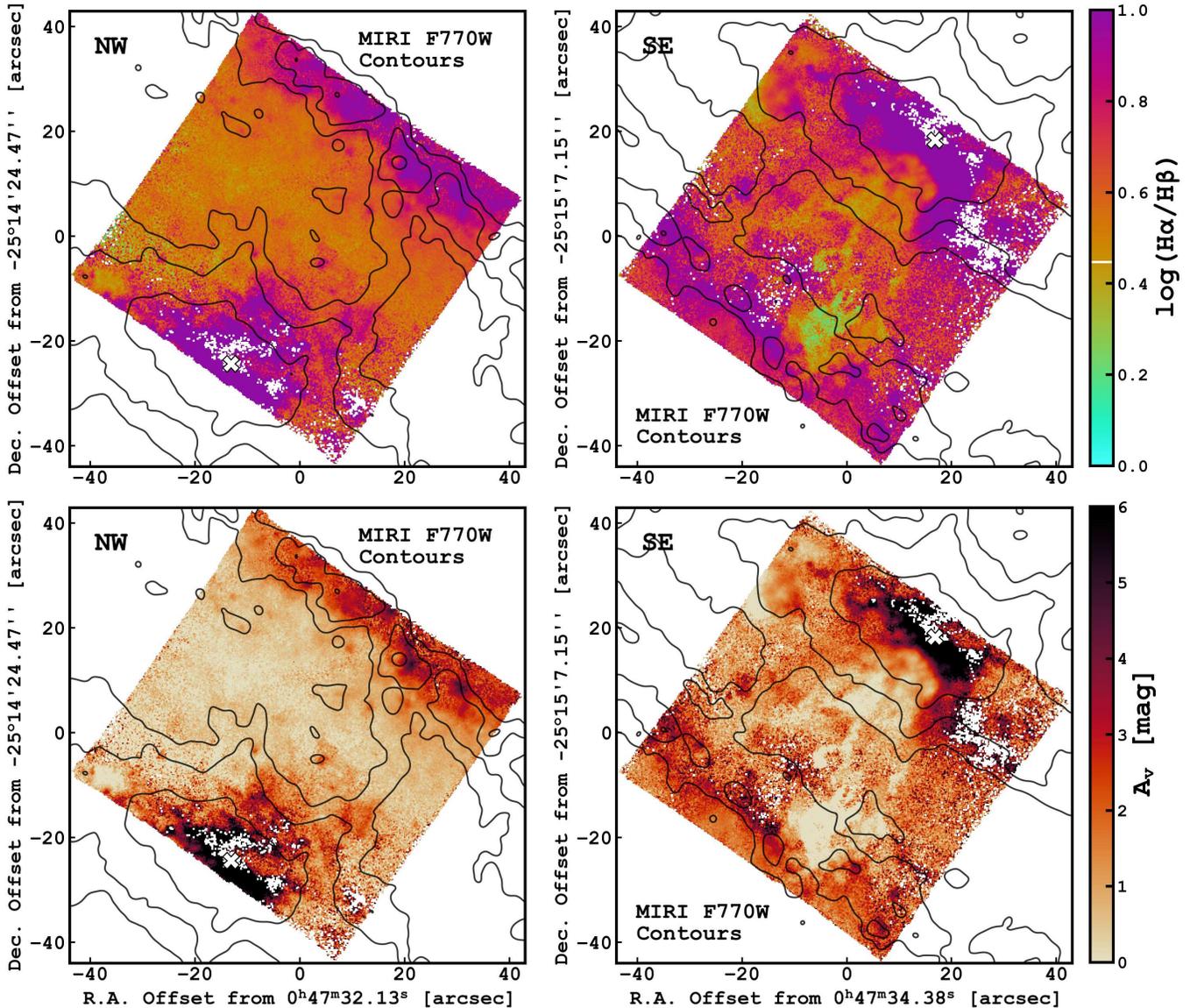

**Figure 12.** Top: Hα/Hβ total integrated flux ratios for the NW (left) and SE (right) cubes. Reddening is signified when Hα/Hβ > 2.8 (0.4 in log-space; white line in the color bar). Bottom: corresponding $A_V$ magnitudes assuming a D. Calzetti et al. (2000) extinction curve, $T_e = 10^4$ K, and $R_V = 3.1$. $A_V$ reaches ≥5 mag in the nucleus and decreases to <1 mag in the outflow. Both: JWST/MIRI F770W contours (levels = [30, 50, 155, 1023] MJy sr$^{-1}$) highlight the nucleus, spiral arms, and outflow filaments. The white × marks the galaxy center.

An important note is that while Case B recombination does not quite hold if the Hα/Hβ ratio is associated with shocks, the intrinsic ratio only increases slightly to Hα/Hβ = 3.0. Changing this value in Equation (2) does not change our Figure 12 results. Thus, shocks have a negligible effect on our inferred $A_V$ values. We further explore shocks and their effects on observed line ratios in the next section.

### 5.4. Emission-line Ratios

Nebular line ratios offer unique diagnoses of the physical conditions of the ISM (D. E. Osterbrock & G. J. Ferland 2006; B. T. Draine 2011). In this section, we use line ratios to determine the dominant source of ionization and electron densities in the southern wind of NGC 253.

#### 5.4.1. Ionization Source

We expect [N II] and [S II] to be brightest (and thus [N II]/Hα and [S II]/Hα to be largest) in the outflow, where collisional and shock ionization should dominate over photoionization (S. Veilleux & D. E. Osterbrock 1987; S. Veilleux & D. S. Rupke 2002; R. G. Sharp & J. Bland-Hawthorn 2010). We present these emission-line ratios for the SE cube in Figure 13. For the [S II]/Hα ratio, we combine the [S II] doublet to increase the S/N. For both ratios, additional blacked-out spaxels indicate low S/N in either line.

The thresholds for where shocks are the likely ionization source are [N II]/Hα > 0.63 and [S II]/Hα > 0.4. For both indicators, we find higher ratio values in the outflow (mean Blue Outflow [N II]/Hα = 1.3 and [S II]/Hα = 1.6) than the disk (mean [N II]/H$\alpha$ = 0.8 and [S II]/H$\alpha$ = 0.4). There are regions in the Disk maps where the [S II]/H$\alpha$ and [N II]/H$\alpha$ ratios fall above the thresholds for nonphotoionization processes. M. S. Westmoquette et al. (2011) also report similar findings. These ratios largely occur in the area where the outflow is projected against the disk, and appear to correspond with broad velocity dispersions (Figure 8). These broad components and line ratios may be the product of disk turbulence and shocks induced by the outflow, or alternatively are regions where the large line width causes blending and misclassifications of components.

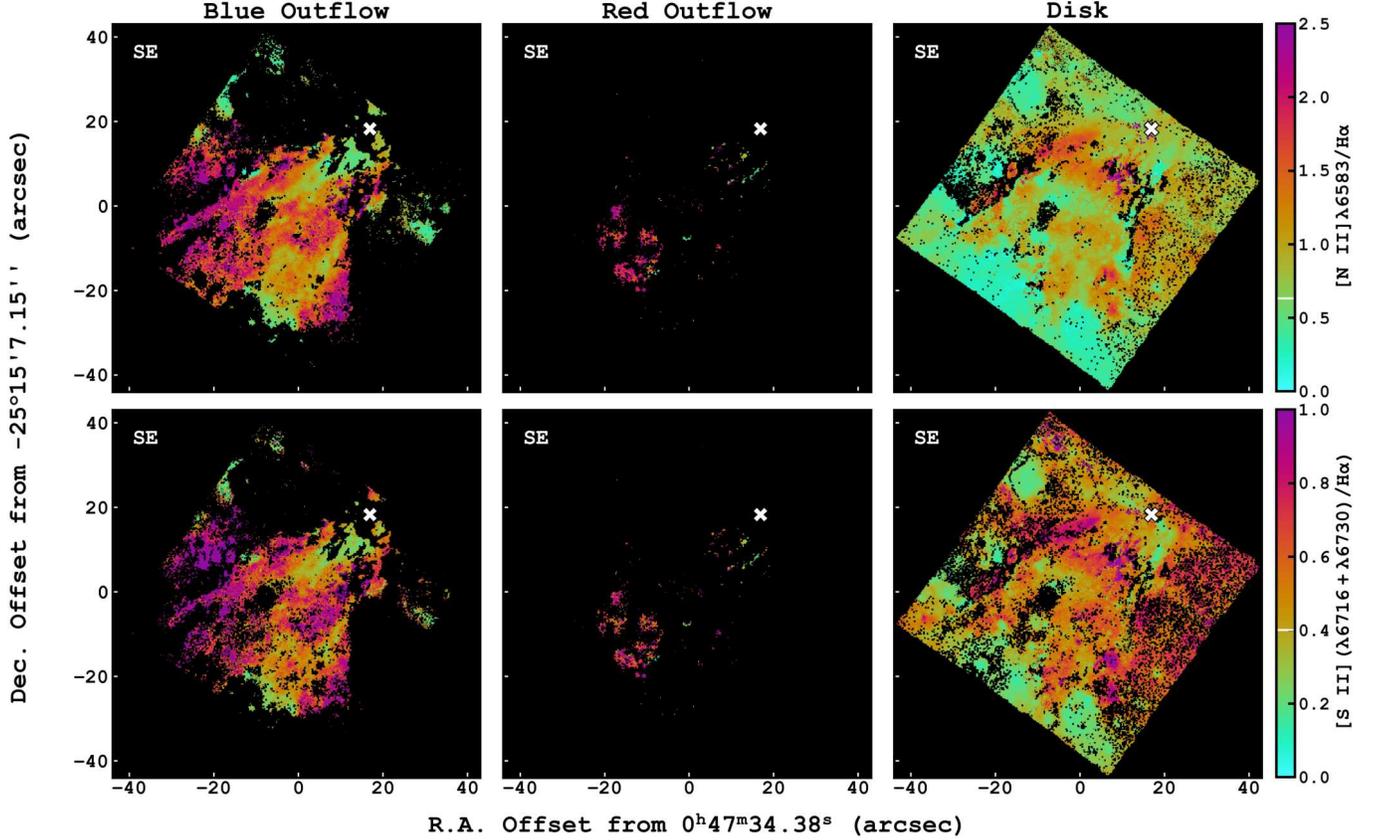

**Figure 13.** [N II]$\lambda$6583/H$\alpha$ and ([S II]$\lambda$6716+[S II]$\lambda$6730)/H$\alpha$ (henceforth, [N II]/H$\alpha$ and [S II]/H$\alpha$) integrated flux line ratios. Both ratios are higher in the extended edges of the outflow cone in comparison with the disk. Both [N II]/H$\alpha$ > 0.63 and [S II]/H$\alpha$ > 0.4 (white line in color bars) are strong indicators of shock ionization. Additional blacked-out spaxels indicate low S/N in either line, and the white × marks the galaxy center.

The edges of the outflow cone are particularly bright in [N II]$\lambda$6583 and [S II], especially in regions adjacent to the SWS and SES. The lowest [N II]/H$\alpha$ and [S II]/H$\alpha$ values are located near the bar and in the spiral arm, coinciding with the strongest H$\alpha$ intensity presented in Figure 6. Both ratios overall appear to be increasing with distance from the nucleus. These findings are in overwhelming agreement with previous studies (K. Matsubayashi et al. 2009; M. S. Westmoquette et al. 2011). Finally, it appears that where the [N II]/H$\alpha$ ratio is high, so is [S II]/H$\alpha$. This suggests a common ionization source (M. A. Dopita & R. S. Sutherland 1995). We note that some of the elevated [S II] values may be biased due to noise since the [S II] doublet is very faint.

The observed brightness of the collisionally excited [N II] and [S II] relative to H$\alpha$ is not easily explained by photoionization alone. Without a known active galactic nucleus (AGN), mechanical heating due to shocks is the likely cause for elevated [N II]/H$\alpha$ and [S II]/H$\alpha$ ratios in NGC 253 (S. Veilleux & D. E. Osterbrock 1987; M. A. Dopita & R. S. Sutherland 1995). The simplest picture is that of a fast-moving wind slamming into the ambient ISM, generating these shocks. Shock speeds upwards of 500 km s$^{-1}$ are needed to explain the highest ratios (M. A. Dopita & R. S. Sutherland 1995). Recent results also suggest that cosmic rays may play a significant role in moving some of the observed line ratios to the AGN region of Baldwin–Phillips–Terlevich diagrams (E. Koutsoumpou et al. 2025).

We compare our predicted line ratios with wind-shock-only models presented in M. G. Allen et al. (2008), valid for fast radiative shocks and $n_e$ = 10–1000 cm$^{-3}$. Our median Blue Outflow [N II]/H$\alpha$ = 1.3, which according to M. G. Allen et al. (2008) is near the ratio value 1.4 predicted by shock velocities around 300 km s$^{-1}$. A median [S II]/H$\alpha$ = 0.63 is predicted by shock velocities near 200 km s$^{-1}$ (the actual ratio being 0.66). Thus, our observed [N II]/H$\alpha$ and [S II]/H$\alpha$ ratios can be predicted by fast shocks up to a few hundred kilometers per second.

*5.4.2. Electron Density*

The [S II]$\lambda$6716/[S II]$\lambda$6730 emission-line flux ratio (hereafter the [S II] ratio) is a diagnostic of electron density ($n_e$). This is because the relative populations of two transitions of the same ion will depend on their collisional excitation rates, which are sensitive to density and only weakly dependent on electron temperature (D. E. Osterbrock & G. J. Ferland 2006; B. T. Draine 2011).

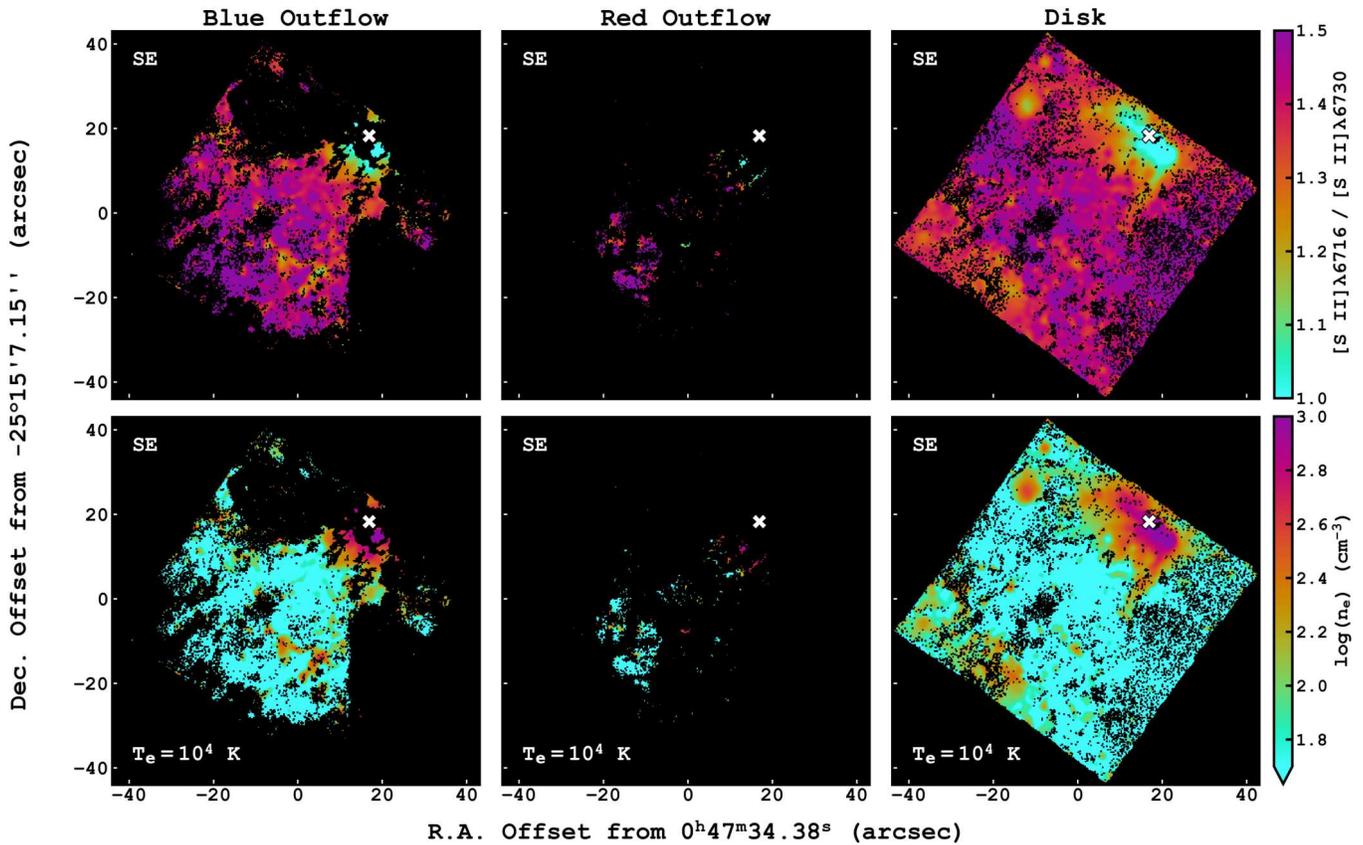

**Figure 14.** [S II]λ6716/[S II]λ6730 smoothed integrated flux line ratio (top) and corresponding electron density $n_e$ (bottom), the latter calculated using the software PyNeb. We assume $T_e = 10^4$ K for an ionized medium. We check against $T_e = 5000$ K and find that $n_e$ values do not change significantly. $n_e$ peaks at ∼2100 cm$^{-3}$ near the base of the outflow, and decreases with distance out to 250 pc, where the wind reaches the low-density regime ($n_e \lesssim 50$ cm$^{-3}$ at $T_e = 10^4$ K; see arrow on color bar). The white × marks the galaxy center.

We present the [S II] ratio in the top panel of Figure 14. Because of low S/N throughout the wind, we smooth the ratio with a Gaussian kernel (standard deviation of 3 pixels). Higher [S II] ratios indicate lower $n_e$, with a theoretical maximum value of ≈1.4 (D. E. Osterbrock & G. J. Ferland 2006). The outflow cone is in the low-density limit away from the wind-launching site.

We convert the [S II] ratio into actual electron densities using the software PyNeb (V. Luridiana et al. 2015), which determines the level populations by solving the equations of statistical equilibrium for an $n$-level atom. For a change of ionized gas temperature of $T_e = 5000$–$10^4$ K, the $n_e$ drops by a factor of 1.4. We present the $T_e = 10^4$ K calculations for $n_e$ in the bottom panels of Figure 14. At this temperature the low-density regime is $n_e \lesssim 50$ cm$^{-3}$.

Areas nearest the nucleus of NGC 253 exhibit $n_e \approx 2100$ cm$^{-3}$, consistent with previous results for the ionized gas density in the nucleus (C. W. Engelbracht et al. 1998; A. Beck et al. 2022). This value is also in line with the MAGNUM survey of nearby galaxies hosting outflows, whose nuclei peak at $n_e \sim 10^3$ cm$^{-3}$ (M. Mingozzi et al. 2019), and the central densities measured in the winds of nearby far-infrared-bright galaxies ($n_e = 500$–$1000$ cm$^{-3}$; T. M. Heckman et al. 1990). The Blue Outflow $n_e$ decreases with distance before reaching the low-density limit about 250 pc from the nucleus. This result is lower than the $n_e \approx 300$ cm$^{-3}$ recorded by M. S. Westmoquette et al. (2011) at 200 pc.

### 5.5. Energetics of the Southern Outflow

With our kinematics and flux measurements, we may estimate the energetics of the southern NGC 253 wind. We focus on Hα gas to avoid having to account for metallicity or the ionization parameter. Throughout this section, we use the Blue Outflow measurements for the southern outflow.

#### 5.5.1. Outflow Mass

The total Hα flux in the southern outflow cone (top-left panel of Figure 6) is $F_{H\alpha} = 9 \times 10^{-13}$ erg cm$^{-2}$ s$^{-1}$, resulting in a luminosity of $L_{H\alpha} = 1.3 \times 10^{39}$ erg s$^{-1}$ ($d = 3.5$ Mpc). From S. Veilleux et al. (2020), the mass of the Hα-emitting outflow is

$$M_{\text{out}} = 3.3 \times 10^8 \frac{C_e L_{H\alpha,44}}{n_{e,3}} M_\odot. \quad (3)$$

This equation is valid for $T_e = 10^4$ K. $C_e \equiv \langle n_e^2 \rangle / \langle n_e \rangle^2$ is the electron density clumping factor (order unity means uniform density). $L_{H\alpha,44}$ is the Hα luminosity normalized to $10^{44}$ erg s$^{-1}$, and $n_{e,3}$ is the mean electron density normalized to $10^3$ cm$^{-3}$.

The electron density derived from the [S II] doublet is sensitive to the ratio between the mean density and $C_e$, whereas $n_e$ in Equation (3) is only the mean density. This means that we cannot use the $n_e$ from the [S II] doublet in this formalism. Instead, we use the following relation from

S. Lopez et al. (2025) to derive average densities from Hα emission, assuming Case B recombination and an optically thin medium:

$$n_e \simeq 10.9 \text{ cm}^{-3} \left( \frac{I_{H\alpha}}{10^{-5} \text{ erg cm}^{-2} \text{ s}^{-1} \text{ sr}^{-1}} \right)^{1/2} \left( \frac{\text{pc}}{\Delta s} \right)^{1/2}. \quad (4)$$

$\Delta s$ is the depth of the medium. Because $I_{H\alpha} \propto n_e^2$, clumpy (high-density) regions will dominate the emission and bias toward higher mean densities (see Section 2 of S. Lopez et al. 2025). We calculate an average Hα intensity from Figure 6 and assume $\Delta s = 40$ pc, the thickness of the outflow cone wall found by M. S. Westmoquette et al. (2011). This yields a mean $n_e \sim 2$ cm$^{-3}$ (well within the low-density limit of the [S II] indicator).

In practice, the clumping factor $C_e$ is a difficult quantity to estimate from the data. We gauge a rough value by analyzing the distribution of high-brightness areas in the outflow (Figure 6). This method determines that high-density structures approximately constitute ~20% of the total Hα-emitting volume (assuming it is a thin wall on the cone surface). This estimate is highly uncertain, and so we test a range of $C_e$. We find that $C_e = 0.1$–0.4 results in reasonable mass and energetics estimates. Therefore, we assume a median $C_e \sim 0.2$. This range of clumpiness is consistent with recent studies of M82, where ionized emission resides in structures ranging from ~0.1 to 100 pc (X. Xu et al. 2023; S. Lopez et al. 2025). Inputting our estimates into Equation (3), we obtain $M_{out} \sim 4 \times 10^5 \, M_\odot$, with significant uncertainty (likely 0.3 dex) due to the effects of clumping.

### 5.5.2. Outflow Energy, Momentum, and Rates

While the bulk of the momentum is thought to be carried by the cold molecular phase (e.g., S. Veilleux et al. 2020), we are solely focused on characterizing the momentum and energetics of the Hα-emitting ionized phase of the NGC 253 wind. To compute the energetics of the Hα-emitting wind, we must assume an average deprojected velocity and cone length. In Section 5.1, we found that $v_{proj} \sim -275$ km s$^{-1}$, typical of velocities at the center of the Blue Outflow map, leads to $v_{deproj} \sim -400$ km s$^{-1}$. We adopt this deprojected velocity for the following calculations. The projected extent of the outflow cone is about 800 pc, so we take an average deprojected $R_{out} \approx 400/\sin(78°)$.

We determine the mass-outflow rate $\dot{M}_{out}$ of the southern outflow to be $\dot{M}_{out} = M_{out} v_{out}/R_{out} \sim 0.4 \, M_\odot \text{ yr}^{-1}$. With a SFR = 2.8 $M_\odot$ yr$^{-1}$, the ionized mass-loading factor is $\eta = \dot{M}_{out}/\text{SFR} \sim 0.1$. This is less than the molecular mass-loading factor ($\eta \approx 8$–20 for $R \lesssim 300$ pc; N. Krieger et al. 2019), which is expected, as the bulk of the outflow mass resides in the cold molecular and neutral phases (A. D. Bolatto et al. 2013; A. K. Leroy et al. 2015; N. Krieger et al. 2019). Our $\eta$ is on the same order as $\eta = 0.37$ for M82 (X. Xu et al. 2023). Additionally, a small survey of photoionized outflows found the relation between the ionized mass-loading factor $\eta$ and $M_\star$ (the mass of a galaxy) to be $\eta \approx 0.76(M_\star/10^{10} \, M_\odot)^{-0.43}$ (J. Chisholm et al. 2017). For an NGC 253 mass of $8 \times 10^{11} \, M_\odot$ (I. D. Karachentsev et al. 2021), we calculate $\eta = 0.1$, consistent with our result. Similarly, the MaNGA survey found a relationship between $\eta$ and the SFR in Hα-emitting regions, which would also put $\eta \approx 0.1$ at SFR = 2.8 $M_\odot$ yr$^{-1}$ (B. Rodríguez del Pino et al. 2019). In summary, our $\eta \sim 0.1$ is in line with expected values.

The average kinetic energy of the Hα-emitting outflow is $E_{out} = 0.5 M_{out} v_{out}^2 \sim 6 \times 10^{53}$ erg. The outflow power is then $\dot{E}_{out} = 0.5 \dot{M}_{out} v_{out}^2 \sim 6 \times 10^{47}$ erg yr$^{-1}$. According to N. Murray et al. (2005) and J. Chisholm et al. (2017), the energy deposition rate due to the starburst (assuming supernovae supply the total energy) is

$$\dot{E}_{SN} = 3 \times 10^{41} \frac{\text{SFR}}{1 M_\odot \text{ yr}^{-1}} \text{ erg s}^{-1}. \quad (5)$$

In the case of NGC 253, $\dot{E}_{SN} = 2.7 \times 10^{49}$ erg yr$^{-1}$ for SFR = 2.8 $M_\odot$ yr$^{-1}$. Thus, the energy efficiency of the southern wind is a mere $\dot{E}_{out}/\dot{E}_{SN} \sim 2\%$. This efficiency is within the range of efficiencies previously calculated for nearby star-forming galaxies with ionized outflows (0.9%–21%; J. Chisholm et al. 2017). We note that the true energy efficiency of the southern outflow may be even lower, as $\dot{E}_{SN}$ only accounts for the initial energy supplied by the starburst and not from other sources.

Similarly, we estimate the outflow momentum and momentum rates to be $p_{out} = M_{out} v_{out} \sim 3 \times 10^{46}$ cm g s$^{-1}$ and $\dot{p}_{out} = \dot{M}_{out} v_{out} \sim 1 \times 10^{33}$ dyn. Following the same assumptions as Equation (5), the momentum deposition rate due to a starburst is

$$\dot{p}_{SN} = 2 \times 10^{33} \frac{\text{SFR}}{1 \, M_\odot \text{ yr}^{-1}} \text{ dyn}. \quad (6)$$

For NGC 253, $\dot{p}_{SN} = 5.6 \times 10^{33}$ dyn, and thus $\dot{p}_{out}/\dot{p}_{SN} \sim 0.2$ for the southern outflow. $\dot{p}_{SN}$ only accounts for the initial starburst (mainly due to supernovae) and does not include other sources of momentum (e.g., the Sedov–Taylor phase). To include all sources of momentum, we scale the initial momentum rate $\dot{p}_{SN}$: $\dot{p}_{SN,final} = 1.3 \, \dot{p}_{SN}$, assuming a mean number density of $n_0 = 1$ cm$^{-3}$ (C.-G. Kim & E. C. Ostriker 2015). This relation results from simulations that assume an inhomogeneous ISM, accounting for the propagation of supernova remnants beyond the initial explosion. Putting it all together, we find $\dot{p}_{out}/\dot{p}_{SN,final} \sim 0.1$. When the ratio between the outflow momentum and the starburst momentum is <1, the momentum imparted by the starburst is sufficient enough to drive the outflow without contributions from another energy source (e.g., an AGN). This is consistent with the current picture of NGC 253 being a purely starburst-driven outflow. We note that our definitions of $\dot{M}_{out}$, $\dot{E}_{out}$, and $\dot{p}_{out}$ imply that these quantities are time-averaged over the dynamical timescale of the outflow: $t_{dyn,out} \sim R_{out}/v_{out} \sim 1$ Myr.

Again, we emphasize that these estimates are highly uncertain (0.3 dex) due to clumping. Additionally, these numbers only account for the southern lobe of the outflow. If we assume the northern lobe is of a similar mass to the southern lobe, then the mass increases by a factor of 2. Even with a range of clumping factors, the *total* outflow energetics remains consistent with $\dot{p}_{out}/\dot{p}_{SN,final} < 1$.

Finally, we must note that our mass estimate of the southern Hα-emitting outflow is lower than the mass found by M. S. Westmoquette et al. (2011). The authors used kinematic modeling and calculated $M_{out} = V \rho f$, where $V$ is the volume of the outflow modeled as a conic frustum, $\rho$ is the density, which is the mass of a proton $m_p$ multiplied by the electron density $n_e$ (which they set to $n_e = 100\,\text{cm}^{-3}$), and $f = 0.1$ is the filling factor. However, we find that their calculated $M_{out} = 6 \times 10^6\,M_\odot$ leads to an overestimate of the energetics; specifically, a mass-outflow rate higher than what has been found for M82 (X. Xu et al. 2023) and a $\dot{p}_{out}/\dot{p}_{SN,final} \sim 2$. While our own mass estimate is uncertain, we believe that the differences between our calculations and those of M. S. Westmoquette et al. (2011) are the assumed electron density and filling factor. Our derived electron density corresponds to a filling factor of $f \sim 0.01$, which is in line with the highly filamentary nature of the wind and diffuse ionized gas in the Milky Way (E. M. Berkhuijsen et al. 2006). Following the steps of M. S. Westmoquette et al. (2011), but instead substituting $n_e \sim 2\,\text{cm}^{-3}$ (and thus $f \sim 0.01$), the mass and energetics converge with the results of this work.

## 6. Summary

In this work, we analyze the kinematics and physical conditions of the NGC 253 superwind by fitting multiple systems of Gaussian profiles to MUSE IFS data cubes. Most spectra can be fit with one and two systems of lines, where multiple components indicate velocity contributions from the outflow in addition to the disk. In summary, our results can be interpreted as follows.

1. Typical projected velocities in the approaching southern outflow cone are between $-150$ and $-250\,\text{km s}^{-1}$. The bluest part of the cone, assuming it is the region most projected toward the observer, has a maximum projected velocity of $\sim -350\,\text{km s}^{-1}$, or a deprojected speed of $\sim -500\,\text{km s}^{-1}$. The kinematics are in line with a biconical outflow with limb-brightened edges due to projection effects.
2. The ionized gas is moving faster with increasing projected distance from the nucleus. On the blue side (negative velocities), the observed projected gradient is $\sim -350\,\text{km s}^{-1}\,\text{kpc}^{-1}$ down the center of the cone and $\sim -550\,\text{km s}^{-1}\,\text{kpc}^{-1}$ along the edge of the cone.
3. Throughout the outflow, the [N II]/H$\alpha$ and [S II]/H$\alpha$ ratios exceed nominal thresholds where shock ionization begins to dominate over photoionization from the nuclear star clusters. The highest line ratios are at the edges of the cone, and tend to increase with distance from the nucleus. Models of fast radiative shocks suggest that velocities of $\sim 200$–$300\,\text{km s}^{-1}$ can account for the observed [N II]/H$\alpha$ and [S II]/H$\alpha$ ratios.
4. Assuming $T_e \approx 10^4\,\text{K}$, electron densities peak around $2100\,\text{cm}^{-3}$ in the nucleus and quickly reach $\lesssim 50\,\text{cm}^{-3}$ in the wind. This value is the low-density limit for our [S II] tracer.
5. The mass of the southern lobe of the H$\alpha$-emitting outflow is $M_{out} \sim 4 \times 10^5\,M_\odot$ with an uncertainty of 0.3 dex. This yields an outflow rate of $\dot{M}_{out} \sim 0.4\,M_\odot\,\text{yr}^{-1}$. The mass-loading factor is then $\eta \sim 0.1$, and the energetic efficiency of the starburst is only $\sim 2\%$. If the northern lobe of the outflow is of a similar mass, these estimates increase by a factor of 2 for the total outflow. These energetics are consistent with a wind solely driven by a starburst.

## Acknowledgments

The authors graciously thank the anonymous reviewer for their very timely and helpful report that improved this work. S.A.C. would like to thank Francesco Belfiore, Ramsey Karim, Emma Mirizio, Jerome Seebeck, and Andrew Harris for helpful discussions. S.A.C., A.D.B., and L.N. acknowledge support from the NSF under award AST-2108140. K.D. acknowledges support from the National Science Foundation Graduate Research Fellowship under grant No. DGE-2236417. K.K. gratefully acknowledges funding from the Deutsche Forschungsgemeinschaft (DFG, German Research Foundation) in the form of an Emmy Noether Research Group (grant No. KR4598/2-1; PI: Kreckel) and the European Research Council's starting grant ERC StG-101077573 ("ISM-METALS").

This work is based on observations collected at the European Southern Observatory under ESO program 0102.B-0078. We acknowledge the usage of the HyperLeda database (http://leda.univ-lyon1.fr). This research has made use of the NASA/IPAC Extragalactic Database (NED), which is operated by the Jet Propulsion Laboratory, California Institute of Technology, under contract with the National Aeronautics and Space Administration. This research has made use of NASA's Astrophysics Data System Bibliographic Services. This work is based on observations made with the NASA/ESA/CSA JWST. The data were obtained from the Mikulski Archive for Space Telescopes at the Space Telescope Science Institute, which is operated by the Association of Universities for Research in Astronomy, Inc., under NASA contract NAS 5-03127 for JWST. These observations are associated with program JWST-GO-01701. Support for program JWST-GO-01701 is provided by NASA through a grant from the Space Telescope Science Institute, which is operated by the Association of Universities for Research in Astronomy, Inc., under NASA contract NAS 5-03127. The specific observations analyzed can be accessed via doi:10.17909/qpdy-1n82.

*Facilities:* VLT:Yepun, ALMA.

*Software:* astropy (Astropy Collaboration et al. 2013, 2018), pyspeckit (A. Ginsburg & J. Mirocha 2011; A. Ginsburg et al. 2022), spectral-cube (T. Robitaille et al. 2016; A. Ginsburg et al. 2019), NumPy (S. Van Der Walt et al. 2011; C. R. Harris et al. 2020), Matplotlib (J. D. Hunter 2007), Pandas (W. McKinney et al. 2010), SciPy (P. Virtanen et al. 2020), Jupyter (T. Kluyver et al. 2016), cmasher (E. van der Velden 2020).

## ORCID iDs

Serena A. Cronin https://orcid.org/0000-0002-9511-1330
Alberto D. Bolatto https://orcid.org/0000-0002-5480-5686
Enrico Congiu https://orcid.org/0000-0002-8549-4083
Keaton Donaghue https://orcid.org/0009-0004-5807-9142
Kathryn Kreckel https://orcid.org/0000-0001-6551-3091
Adam K. Leroy https://orcid.org/0000-0002-2545-1700
Rebecca C. Levy https://orcid.org/0000-0003-2508-2586
Sylvain Veilleux https://orcid.org/0000-0002-3158-6820
Fabian Walter https://orcid.org/0000-0003-4793-7880